\documentclass[12pt]{article}

\usepackage{amsmath}
\usepackage{amssymb}
\usepackage{graphicx} 

\advance\hoffset by -7mm   
\makeatletter 
\@addtoreset{equation}{section}  
\makeatother 
 
\newcommand{\ii}{\'\i}

\newcommand{\cao}{\c c\~ao}

\newcommand{\ftoday}{{\sl {Le \number\day \space\ifcase\month 
\or janvier\or f\'evrier\or mars\or avril\or mai
\or juin\or juillet\or ao\^ut\or septembre\or octobre
\or novembre \or d\'ecembre\fi\space \number\year}}}    
\newcommand{\ptoday}{{\sl {\number\day \space de\space \ifcase\month 
\or janeiro\or fevereiro\or mar{\c c}o\or abril\or maio
\or junho\or julho\or agosto\or setembro\or outubro
\or novembro \or dezembro\fi\space de\space \number\year}}}    
\newcommand{\gtoday}{{\sl {Den \number\day. \ifcase\month 
\or Januar\or Februar\or M\"arz\or April\or Mai
\or Juni\or Juli\or August\or September\or Oktober
\or November \or Dezember\fi\space \number\year}}}    
\newcommand{\journal}[4]{{\em #1~}#2\,(#3)\,#4}

\newcommand{\pr}{\journal {Phys. Rev.}}

\newcommand{\prl}{\journal {Phys. Rev. Lett.}}
\newcommand{\jmp}{\journal {J. Math. Phys.}}

\newcommand{\cqg}{\journal {Class. Quantum Grav.}}

\newcommand{\np}{\journal {Nucl. Phys.}}
\newcommand{\pl}{\journal {Phys. Lett.}}

\newcommand{\prep}{\journal {Phys. Rep.}}

\renewcommand{\a}{\alpha}
\renewcommand{\b}{\beta}
\newcommand{\g}{\gamma}           
\renewcommand{\d}{\delta}         
\newcommand{\e}{\epsilon}
\newcommand{\ve}{\varepsilon}

\newcommand{\la}{\lambda}        \newcommand{\LA}{\Lambda}
\newcommand{\m}{\mu}
\newcommand{\n}{\nu}
\newcommand{\om}{\omega}         \newcommand{\OM}{\Omega}
\newcommand{\s}{\sigma}           
         
           \newcommand{\F}{{\Phi}}
\newcommand{\vf}{{\varphi}}

\newcommand{\LL}{{\cal L}}
\newcommand{\MM}{{\cal M}}

\newcommand{\TT}{{\cal T}}

\newcommand{\esp}{\\[3mm]}

\newcommand{\sla}{\raise.15ex\hbox{$/$}\kern -.57em} 
\newcommand{\Sla}{\raise.15ex\hbox{$/$}\kern -.70em}
\newcommand{\h}{\hbar}
\newcommand{\Lp}{\displaystyle{\biggl(}}
\newcommand{\Rp}{\displaystyle{\biggr)}}

\newcommand{\lp}{\left(}\newcommand{\rp}{\right)}
\newcommand{\lc}{\left[}\newcommand{\rc}{\right]}

\newcommand{\complex}{{\kern .1em {\raise .47ex
\hbox {$\scriptscriptstyle |$}}
    \kern -.4em {\rm C}}}
\newcommand{\real}{{{\rm I} \kern -.19em {\rm R}}}
\newcommand{\rational}{{\kern .1em {\raise .47ex
\hbox{$\scripscriptstyle |$}}
    \kern -.35em {\rm Q}}}
\renewcommand{\natural}{{\vrule height 1.6ex width
.05em depth 0ex \kern -.35em {\rm N}}}

\newcommand{\half}{\frac{1}{2}}

\newcommand{\dfud}[2]{{\displaystyle{\frac{\delta #1}{\delta #2}}}}
\renewcommand{\dfrac}[2]{{\displaystyle{\frac{#1}{#2}}}}
   
\newcommand{\dint}{\displaystyle{\int}}
\newcommand{\eg}{{\em e.g.,\ }}
\newcommand{\Eg}{{\em E.g.,\ }}
\newcommand{\ie}{{{\em i.e.},\ }}

\newcommand{\etc}{{\em etc.\ }}

\newcommand{\twiddle}{\lower.9ex\rlap{$\kern -.1em\scriptstyle\sim$}}

\newcommand{\equ}[1]{(\ref{#1})}
\newcommand{\eq}{\begin{equation}}
\newcommand{\eqn}[1]{\label{#1}\end{equation}}
\newcommand{\eea}{\end{eqnarray}}
\newcommand{\eqa}{\begin{eqnarray}}
\newcommand{\eqan}[1]{\label{#1}\end{eqnarray}}
\newcommand{\ba}{\begin{array}}
\newcommand{\ea}{\end{array}}
\newcommand{\eqac}{\begin{equation}\begin{array}{rcl}}
\newcommand{\eqacn}[1]{\end{array}\label{#1}\end{equation}}

\newcommand{\bz}{\begin{enumerate}}
\newcommand{\ez}{\end{enumerate}}

\newcommand{\ADS}{(A)dS}
\newcommand{\ads}{(a)ds}



\begin{document}

\title{A Topological-like Model for Gravity\\ in 4D Space-time}

\author{Ivan Morales
, 
Bruno Neves, 
Zui Oporto
and Olivier Piguet
\\[4mm]
{\small Departamento de F\ii sica, 
Universidade Federal de Vi\c cosa -- UFV}\\{\small 
 Vi\c cosa, MG, Brazil}}
   
\date{}

\maketitle

\begin{center} 

\vspace{-5mm}

{\small\tt E-mails:
 mblivan@gmail.com, bruno.lqg@gmail.com, \\ azurnasirpal@gmail.com,
opiguet@pq.cnpq.br\esp
Keywords:
Gravitation, Topological theories, Cosmology, Quantum gravity}

\end{center}


\begin{abstract}In this paper we consider a model for gravity in 
4-dimensional space-time originally proposed by Chamseddine, 
  which may be  derived by  dimensional reduction and truncation from a 
5-dimensional Chern-Simons theory. Its topological origin makes it an interesting 
candidate for an easier quantization, \eg in the Loop Quantization framework.
The present paper is dedicated to a classical analysis of the model's properties. 
Cosmological solutions as well as   
wave solutions are found and compared with the corresponding solutions 
of Einstein's General Relativity with cosmological constant.

\end{abstract}

\section{Introduction}

Gravitation as described by Einstein's General Relativity is notoriously difficult to 
reconcile with Quantum Theory, a task which is nevertheless necessary
 if one want to  understand physics at the very small scale defined by the 
 Planck length $l_{\rm P}\sim 10^{-35}\,$m, with also the hope that 
 quantum mechanics will cure the singularities of the classical theory 
 such as the Big Bang and 
 Black Hole ones. Loop Quantum Gravity~\cite{Rovelli,Thiemann,Rovelli-Vidotto} 
 (LQG) is one attempt to do it. It starts from Einstein's classical General Relativity 
 (GR) in the Ashtekar-Barbero formalism where the dynamical variables are an 
 SU(2) Yang-Mills type connection together with its canonical conjugate 
 momentum field. The dynamics is expressed as a set of constraints which 
 correspond to the gauge invariances of the theory~\cite{Dirac-etal}, 
 namely SU(2) local invariance and the invariances under the space and 
 time diffeomorphisms. The quantum theory is then defined by constructing 
 a  Hilbert space whose elements are certain wave functionals of the 
 connection obeying the constraints. The latter should be 
well defined as 
self-adjoint operators, and then solved in the sense that they select the physical wave functionals as those which are annihilated by them. The main difficulty is in the 
definition of the time diffeomorphism constraint and its 
solution. Much progress,   
has been 
made by Thiemann and collaborators~\cite{Thiemann}, and more recently by
Rovelli and collaborators in the spin foam formalism~\cite{Rovelli-Vidotto},
leading to very promising results.

GR is a ``background invariant theory'', which means that no 
a-priory geometric structure is given to the space-time 
manifold where the theory is defined: the metric belongs 
to the dynamical fields. Another class of background 
independent theories is provided by topological theories 
 such as the Chern-Simons (CS) theories\footnote{We could also mention
the $BF$ theories. However, unless some constraints are 
applied to them, they care of local degrees of freedom in 
any space-time dimension.}. Remarkably enough~\cite{Witten}, 
gravity in 3 space-time dimensions can be written as a CS theory whose gauge group is  
the local Poincar\'e group ISO(1,2),   but also SO(1,3) or SO(2,2) if there is a 
positive or negative  cosmological constant. The question is: could 
one describe higher dimension gravity as a CS 
theory~\cite{Lovelock,Teitel-Zan,higherCS,Zanelli-lecture,Chamseddine,Zanelli} 
An essential difference between gravity in 3D and gravities in more than 3 dimensions is that the former has no local degree of freedom, whereas the latter do have. The same happens for the CS theories in 3D and in more than 3 dimensions. Since CS theories live in odd-dimensional space-times, the first one which admits local degrees of freedom is the one in 5D, with 
the gauge group ISO(1,4), SO(1,5) or SO(2,4) (Poincar\'e, de Sitter 
or anti-de Sitter). 

An advantage of topological theories -- 
with the gauge groups mentioned above -- is that (some of) the diffeomorphism
invariances are consequences of the gauge invariances. For CS in 3D,
 all diffeomorphism invariances follow   on shell  from gauge 
invariance~\cite{Isler}, whereas, in 5D, only the invariance 
under the time diffeomorphisms follows   on shell~\cite{Banados-etal}.  
In the latter case this means that the constraint associated with the time 
diffeomorphisms is a consequence of the other constraints. Thus the
difficult task of defining and solving the diffeomorphism constraint 
in the 5D quantum gravity described by this CS theory would be avoided.

Being interested in 4D gravity, can we find a similar, topological-like theory?
An answer has been provided by Chamseddine~\cite{Chamseddine}: 
 a theory which, beyond containing the gravitation fields is 
also containing a dilaton-type scalar field.   It  can be derived from the 5D CS theory by dimensional reduction and truncation of some of the component fields. 
As we will show, the set of solutions of the Chamseddine model is 
a subset of the solutions of the complete, non-truncated 5D CS 
theory reduced  to 4D. As such, in view of the interesting  
properties concerning the constraints
mentioned above, it is worthwhile to  study the classical aspects of the Chamseddine model, which is the purpose of the present work. A study of the whole 5D CS theory, with or without Kaluza-Klein dimensional reduction, 
classically and quantically, will be reported elsewhere~\cite{futurework}.
More specifically, the proposal of the present work is to investigate 
the dynamics of the Chamseddine model and compare some of its solutions with solutions of the conventional Einstein theory. We will 
in particular focus on solutions of the cosmological type  and  wave solutions. 

The present paper begins in Section \ref{Chamseddine model} with a review of Chamseddine's 
derivation of his 4D model -- whose gauge invariance is de Sitter SO(1,4) or
anti-de Sitter SO(2,3) -- from a SO(1,5) or SO(2,4)
5D Chern-Simons theory by dimensional reduction and truncation of some
fields. We clarify some points of this truncation, and moreover show 
through a well chosen gauge fixing that the model is a theory of a 
dilaton-like scalar field interacting with a gravitational field 
with torsion. 
  We also show here that the field equations of the Chamseddine 
theory are special solutions of the full runtruncated CS theory 
reduced to 4D. 
Linear approximations are studied in section 
\ref{Linear approximations}, leading to the Newtonian limit and to 
gravitational wave solutions.
Section \ref{Cosmological solutions} is devoted to the study of cosmological solutions of the theory and their comparison with the $\LA$CDM model. Conclusions and outlooks are presented in Section 5. Conventions and notations are displayed in an appendix.

\section{The Chamseddine model}\label{Chamseddine model}
\subsection{5D \ADS\ Chern-Simons theory as a theory of gravity in 5D}

We start with a description of the 5-dimensional Chern-Simons (CS) theory
for the (anti-)de Sitter gauge group    and its interpretation as a gravitation 
theory~\cite{higherCS} 
(See~\cite{Zanelli-lecture} for a comprehensive review.). 
Our notations and conventions are summarized in the 
Appendix. 

The gauge group transformations are those which 
leave invariant the metric
$\eta_{MN} =$ diag$(-1,1,1,1,1,s)$, with $M,N,\cdots$ = $0,\cdots,5$ and
where  $s$ takes the values $\pm1$. 
The signatures $(-1,1,1,1,1,1)$ and $(-1,1,1,1,1,-1)$ 
correspond respectively to the Minkowskian de Sitter 
group SO(1,5) and anti-de Sitter group SO(2,4)
for 5D space-time, respectively. 
  They will be collectively denoted\footnote{The suffix 
$N$ in \ADS$_N$  makes explicit  the dimension of 
the  defining representation space.} by \ADS$_6$. 

A basis of the Lie algebra \ads$_6$ of \ADS$_6$ is given by the generators 
$M_{MN}$ = $-M_{NM}$,  realized as the 6$\times$6 matrices
$(M_{MN})^P{}_Q:=$
$-\delta_M^P\eta_{NQ}+\delta_N^P\eta_{MQ}$, obeying the commutation relations
\eq
 [M_{MN}, M_{PQ}] =
\eta_{MP}M_{NQ}-\eta_{MQ}M_{NP}-\eta_{NP}M_{MQ}+\eta_{NQ}M_{MP}\,.
\eqn{so6-alg}

The field variables\footnote{Fields and forms in 5 dimensional space-time
$\MM_5$
are written with a hat, space-time indices being denoted by $\a,\b,\cdots$ = 
$0,\cdots,4$.} of 
the theory are the components of a 
connection form $\hat A$  = $ \hat{A}_\a dx^\a$,
with values in the Lie algebra \ads$_6$. 
In the basis $(M_{MN})$, the connection form reads
\eq
\hat A = \half \hat A^{MN} M_{MN} = \half \hat A^{MN}_\a dx^\a M_{MN}\,,
\eqn{so6-connection}
and transforms as
\eq 
\d \hat A^{MN} = \hat d\hat \e^{MN} + \hat A^M{}_P\hat \e^{PN}-\hat A^N{}_P\hat \e^{PM}
\eqn{ADS6-transf}
under the infinitesimal \ADS$_6$ gauge transformations.
The gauge invariant CS action is given by\footnote{We don't write
explicitly the wedge symbol $\wedge$ for the external products of forms.
\Eg the product $(\hat{A}^2)^{PQ}=-(\hat{A}^2)^{QP}$ stems for 
$\hat{A}^{PT}\wedge \hat{A}_T{}^Q$.}
\eq\ba{l}
S_{\rm CS } =\esp
\dfrac{1}{24} \ve_{MNPQRS}\!\dint_{{\!\!\!}\!\!\!\MM_5}\!\!
\lp \hat{A}^{MN}\hat{d}\hat{A}^{PQ}d\hat{A}^{RS} + 
\dfrac32 \hat{A}^{MN}(\hat{A}^2)^{PQ}d\hat{A}^{RS} 
+ \dfrac35 \hat{A}^{MN}(\hat{A}^2)^{PQ}(\hat{A}^2)^{RS} \rp\,.
\ea\eqn{CS5action} 
$\ve_{MNPQRS}$ is the Levi-Civita totally antisymmetric tensor 
with the normalization condition\footnote{Indices are lowered 
and raised using the metric $\eta_{MN}$. } $\ve_{012345}=1$.
The field equations obtained by varying the connection $\hat{A}$ are
\eq
\dfrac{1}{4} \ve_{MNPQRS}\hat F^{PQ}\hat F^{RS} = 0\,,\quad
\mbox{where}\quad \hat  F^{PQ} = \hat d\hat{A}^{PQ} + (\hat{A}^2)^{PQ}\,,
\eqn{CS5fieldeq}
with ${\hat F}^{MN}=\hat{d}\hat{A}^{MN} + \hat{A}^N{}_P \hat{A}^{PN}$
the Yang-Mills curvature.

In the same way as one can interpret the 3-dimensional CS theory for the pseudo-orthogonal 
gauge group SO(1,3) or SO(2,2) as a gravitation theory 
with cosmological constant~\cite{Witten}, one can indeed do the same here making the following identifications of the generators $M_{MN}$ 
with the 5-dimensional Lorentz generators $M_{AB}$ and ``translation'' generators $P_A$,
with $A,B,\cdots$ = $0,\cdots,4$:
\eq
M_{AB} = M_{AB}\,,\quad P_A := {\la}  M_{A5}\,,
\eqn{so6toso5-gen}
where $\la>0$ is a parameter with the dimension of a mass 
or of the 
inverse of a length in the system of units where $c=\h=1$.
The commutation relations \equ{so6-alg} take the explicit form of the 
\ads$_6$ commutation relations \equ{alg-(A)dS_6->Lorentz}, with 
$\eta_{AB}$ being the 5-dimensional Lorentz metric
$(-1,1,1,1,1)$ and $s|\la^2|$ playing the role of 
the ``cosmological constant'' (See Appendix \ref{A - Lie algebra basis}).
We define accordingly the spin connection form 
$\hat{\om}^{AB}$ and the 5-bein
form $\hat e^A$ as
\eq
\hat \om^{AB} := \hat A^{AB}\,,\quad 
\hat e^A := \frac{1}{\lambda } \hat A^{A5}\,.
\eqn{so6toso5-comp}

We can thus write the infinitesimal \ADS$_6$ gauge 
transformations 
\equ{ADS6-transf} and the CS action \equ{CS5action} 
as\footnote{A subscript will be used to distinguish 
the \ADS$_5$ curvature from
the Lorentzian SO(1,3) curvature. We opt not to write the subscript in the
case of the latter. Thus we have
\begin{align*}
\hat{R}_{(5)}^{AB} & =\hat{ d}\hat{\omega}^{AB}+\hat{\omega}^{A}\!_{C}\hat{\omega}^{CB},\\
\hat{R}^{IJ}& =\hat{ d}\hat{\omega}^{IJ}+\hat{\omega}^{I}\!_{K}\hat{\omega}^{KJ}.
\end{align*}
Let us note that the latter is not just obtained by restricting the
former to the indices $I,J$, instead we have that
\[
\hat{R}_{(5)}^{IJ}=\hat{R}^{IJ}-\hat{\omega}^{4I}\hat{\omega}^{4J}.
\]
}
\eq\ba{l}
\d  \hat\om^{AB} = \hat d\hat \e^{AB} +  \hat\om^A{}_C\hat \e^{CB}- \hat\om^B{}_C\hat \e^{CA}
+\lambda  \lp \hat\e^A{}_5\hat e^B - \hat\e^B{}_5\hat e^A  \rp \,,\esp
\d \hat e^A = \hat e^C\hat\e_C{}^A  +
\dfrac{1}{\lambda } \lp \hat d\hat\e^{A5}+ \hat\om^A{}_C\hat\e^{C5} \rp \,,
\ea\eqn{ADS6/5-transf}
\begin{equation}
S^{(5D)}=\dfrac{\la}{8} \!\int_{\mathcal{M}_{5}}\!\ve_{ABCDE}\!\left(\hat{e}^{A}\hat{R}_{(5)}^{BC}\hat{R}_{(5)}^{DE}\!-\!\frac{2s\la^2}{3}\hat{e}^{A}\hat{e}^{B}\hat{e}^{C}\hat{R}_{(5)}^{DE}\!+\!\frac{\la^4}{5}\hat{e}^{A}\hat{e}^{B}\hat{e}^{C}\hat{e}^{D}\hat{e}^{E}\right).\label{eq:5d-Eins-Hilb}
\end{equation}
We recognize in the second and third terms the standard Einstein-Hilbert
and cosmological terms of 5D gravity, respectively. The novelty of the
Chern-Simons action is the appearance of the first term, which is of the
form $e\hat R_{(5)}\hat R_{(5)}$; this term does not enter through an arbitrary
coupling constant, but instead trough a rational number which is pre-fixed
by the requirement of the theory to be \ADS$_6$ invariant, although  
the action is written in a manifestly   Lorentz SO(1,4)  invariant form. 

Since later on we will proceed to a dimensional reduction from 5D to 4D,
we will need the decomposition of the \ADS$_6$ group in terms of   4D
Lorentz SO(1,3)  representations, as displayed in (\ref{MPQR}, \ref{SO(6)->SO(4)}).
The connection components are accordingly decomposed as
\[
\hat{\omega}^{AB}=\{\hat{\omega}^{IJ},\
\hat{\omega}^{I4}=:\lambda \hat{b}^{I}\},\quad\hat{e}^{A}
=\{\hat{e}^{I},\hat{e}^{4}\}.
\]
 With these definitions,  
the action \eqref{eq:5d-Eins-Hilb} writes
\begin{align}
S^{(5D)} & =  \dfrac{1\la}{8} \!\int_{\mathcal{M}_{5}}\!\ve_{IJKL}
\Lp \hat{e}^{4}
(\hat{R}^{IJ}-\la^2(\hat{b}^{I}\hat{b}^{J}+s\hat{e}^{I}\hat{e}^{J}))
(\hat{R}^{KL}-\la^2(\hat{b}^{K}\hat{b}^{L}+s\hat{e}^{K}\hat{e}^{L}))
\!\nonumber\\
 &    \qquad+2\la\hat{D}\hat{e}^{I}\hat{b}^{J}(\hat{R}^{KL}
 -\frac{2\la^2}{3}\hat{b}
 ^{K}\hat{b}^{L})- 2\la\hat{D}\hat{b}^{I}\hat{e}^{J}
 (\hat{R}^{KL}-\frac{2s\la^2}{3}
 \hat{e}^{K}\hat{e}^{L})\Rp\,,
\label{eq-5d}\end{align}
with $\hat{D}$ the covariant external derivative:
$\hat D \hat e^I=\hat d \hat e^I + \hat\om^I{}_J\hat e^J$, \etc, and
$\hat{R}^{IJ}$ = $d \hat \om^{IJ} + \hat\om^I{}_K\hat\om^{KJ}$.
Let us note that the fields $\hat{e}^{I}$ and $\hat{b}^{I}$ play a symmetrically
role in the action, so in principle we can use any of both to define
a 4-dimensional soldering form. A qualitative difference between these
quantities will show up after a suitable truncation.

In view of the announced dimensional reduction,
we make explicit a split of the D=5 space-time coordinates  in
 D=4 space-time coordinates $x^\m$, $\m=0,\cdots,3$ and 
 the fifth coordinate $\chi:=x^4$ by  writing the form fields as
\eq\ba{l}
\hat{e}^{I}  =e_{\mu}^{I} d x^{\mu}+e_{\chi}^{I} d\chi,\esp
\hat{b}^{I}  =b_{\mu}^{I} d x^{\mu}+b_{\chi}^{I} d\chi,\esp
\hat{e}^{4}  =e_{\mu}^{4} d x^{\mu}+{e^4_\chi} d\chi,\esp
\hat{\omega}^{IJ}  =\omega_{\mu}^{IJ} d x^{\mu}+\omega_{\chi}^{IJ} d\chi.
\ea\eqn{chi-split}
  The corresponding splits for the curvature components read
\eq
\hat{F}^{MN}=F^{MN}+F_{\chi}^{MN} d\chi
\eqn{F-split}
where
\[
F^{MN}=\frac{1}{2}F_{\mu\nu}^{MN} d x^{\mu} d x^{\nu},\quad F_{\chi}^{MN}=F_{\mu\chi}^{MN} d x^{\mu}.
\]
We also have to split each of these forms in terms of their Lorentz SO(1,3)
components,
\[
F^{MN}=(F^{IJ},F^{4I},F^{5I},F^{45}),\quad F_{\chi}^{MN}=(F_{\chi}^{IJ},F_{\chi}^{4I},F_{\chi}^{5I},F_{\chi}^{45}).
\]
Then with the relabelling 
\equ{chi-split}, \equ{F-split},
the curvature components take the form
\eq\ba{l}
F^{IJ}  =R^{IJ}-\la^2b^{I}b^{J}-s\la^2e^{I}e^{J}\,,\esp
F^{I4}  = \la Db^I - s\la^2 e^I e^4   \,,\esp
F^{I5}  = \la D e^{I} + \la^2 b^I e^4    \,,\esp
F^{45}  = \la d e^4 - \la^2 b_I e^I \,,\esp
F_{\chi}^{IJ}  = R_\chi^{IJ}+\la^2(b_\chi^I b^J - b^Ib_\chi^J) 
+s\la^2(e_\chi^I e^J - e^Ie_\chi^J)   \,,\esp 
F_{\chi}^{I4} = \la( D b_\chi^{I} + \om_\chi^{I}{}_J b^J)
+  s\la^2 (e^{I}_\chi e^{4} - e^{I} e_\chi^{4})\,\esp
F_{\chi}^{I5} = \la( D e_\chi^{I} + \om_\chi^{I}{}_J e^J)
- \la^2 (b^{I}_\chi e^{4} - b^{I} e_\chi^{4})    \,,\esp
F_{\chi}^{45} = \la d e_\chi^{4} + \la^2 (b^{I}_\chi e^{I} - b^{I} e_\chi^I)  \,,
\ea\eqn{field-eq-Lorentz-comp}
where $D$, and $R^{IJ}$ represent the covariant exterior derivative
and the curvature 2-form associated 
to the Lorentz connection $\om^{IJ}$.
The field equations  \eqref{CS5fieldeq}  are then splitted into 4-forms equations and 
3-form equations (the $\chi$-components). The 4-form equations are:
\begin{gather}
\begin{split}
\ve_{IJKL}(F^{45}F^{KL}-2F^{K4}F^{L5}) & =0\,,\\
\ve_{IJKL}F^{J5}F^{KL} & =0\,,\\
\ve_{IJKL}F^{J4}F^{KL} & =0\,,\\
\ve_{IJKL}F^{IJ}F^{KL} & =0\,,
\end{split}\label{eq:auxiliar-4}
\end{gather}
and the 3-form equations are:
\begin{gather}
\begin{split}
\ve_{IJKL}(F_{\chi}^{45}F^{KL}+F^{45}F_{\chi}^{KL}
-2F_{\chi}^{K4}F^{L5}-2F^{K4}F_{\chi}^{L5}) & =0\,,\\
\ve_{IJKL}(F_{\chi}^{J5}F^{KL}+F^{J5}F_{\chi}^{KL}) & =0\,,\\
\ve_{IJKL}(F_{\chi}^{J4}F^{KL}+F^{J4}F_{\chi}^{KL}) & =0\,,\\
\ve_{IJKL}F^{IJ}F_{\chi}^{KL} & =0\,,
\end{split}\label{eq:auxiliar-3}
\end{gather}
with the curvature components given by \equ{field-eq-Lorentz-comp}.

This theory is still invariant under the full \ADS$_6$ transformations, which now read, in terms of the 5D ``hat'' quantities:
\eq\ba{l}
\d \hat\om^{IJ} = \hat D\hat \e^{IJ}
+\lambda  s \lp \hat\e^{I5}\hat e^J - \hat\e^{J5}\hat e^I  \rp
+\lambda  \lp \hat\e^{I4}\hat b^J - \hat\e^{J4}\hat b^I  \rp  \,,\esp
\d \hat e^I = \lp 1/\lambda \rp \hat D\hat\e^{I5}
+ \hat e^J\hat\e_J{}^I  + \hat b^I\hat\e^{45} 
- {\hat{e}^4_\chi}\e^{I4} \,,\esp
\d \hat b^I = \lp 1/\lambda \rp \hat D\hat\e^{I4}
+ \hat b^J\hat\e_J{}^I  - s\hat e^I\hat\e^{45} + s\hat e_\chi^4\e^{I5} \,,\esp
\d \hat e^4 = \lp 1/\lambda \rp  \hat d\hat\e^{45}
-\hat b_I\hat\e^{I5} + \hat e_I\hat\e^{I4}\,.
\ea\eqn{ADS6/4-transf}

\subsubsection*{A partial gauge fixing}

The action \equ{eq:5d-Eins-Hilb} or \equ{eq-5d} and the field 
equations (\ref{eq:auxiliar-4}, \ref{eq:auxiliar-3}) may be 
simplified by a partial gauge fixing
consisting in the 8 conditions
\eq
b_\chi^I=0\,,\quad e_\chi^I=0\,,\quad I=0,\cdots,3\,,
\eqn{two-gaugefix}
which fix the gauge symmetries generated by 
$M_{I5}=P_I/{\la}$ and 
$M_{I4}$ = $Q_I/{\la}$,
respectively, as can be inferred from
the transformation laws \equ{ADS6/4-transf} for the $\chi$ components
of $\hat e^I$ and $\hat b^I$, the field $\hat e_\chi^4$ being 
assumed not to vanish. 
This reduces the explicit gauge symmetry to the group
SO(1,3) $\times$ \ADS$_2$, where SO(1,3) is the 4D Lorentz group and 
\ADS$_2$ = U(1) if $s>0$ (theory with positive cosmological constant) or the dilatation group if 
$s<0$ (theory with negative cosmological constant). Of course 
\equ{two-gaugefix} is only a  gauge fixing:
 the theory remains a full \ADS$_6$  gauge theory.

\subsection{The Chamseddine action}

A 4D theory may be obtained trough a Kaluza-Klein dimensional
compactification in which ''matter-like`` fields are realized as  5th dimension components
of the 5D fields.
In our context we may assume that the fifth dimension, of coordinate
 $x^4=\chi$, is compact and ``microscopic'', and the fields are expanded in Kaluza-Klein Modes. 
In the present  paper, we restrict the study to the zero-mode sector -- which amounts to consider all fields as constant in $\chi$ -- leaving a complete discussion involving all modes for future work~\cite{futurework}. This means
\begin{equation}
\partial_{\chi}f(x)=0\,,\quad \forall \ \mbox{field}\ f\,.
\label{eq:zero-mod-cond}
\end{equation}
The Chamseddine model has been obtained~\cite{Chamseddine} 
 by a truncation consisting in setting some fields to zero:
\eq
e_{\chi}^{I} =0\,,\quad \omega_{\chi}^{IJ} = 0\,,\quad
e_{\mu}^{4} =0\,,\quad b_{\mu}^{I}  =0\,. 
\eqn{eq:truncation}
We may observe that the first condition is in fact nothing but a gauge fixing condition, the second of \equ{two-gaugefix}. The other three truncation conditions do indeed break \ADS$_6$, apparently to 
SO(1,3). However, a reordering of the remaining fields in new multiplets allows 
to show that the resulting theory actually has a hidden 
\ADS$_5$ gauge invariance~\cite{Chamseddine}. 
In order to see this, one does not apply for the moment the first of the  gauge fixing conditions
\equ{two-gaugefix}, and reorder the fields in \ADS$_5$ multiplets as
\begin{align}
\mathbb{A}^{AB} & =\{\mathbb{A}^{IJ},\mathbb{A}^{I4}\}
:=\{\omega^{IJ},\lambda e^{I}\},\label{eq:SO5-connec}\\
\Phi^{A} & =\{\Phi^{I},\Phi^{4}\}:=\{-b_{\chi}^{I},e_{\chi}^{4}\}.\label{eq:SO5-Phi}
\end{align}
Using these definitions together with the truncation conditions 
\equ{eq:truncation},
the action \equ{eq:5d-Eins-Hilb} or \equ{eq-5d} reduces to the obviously \ADS$_5$ invariant
expression\footnote{No parameter is needed in the front of the action, since any 
such parameter may be absorbed in a redefinition of the scalar field $\Phi^A$.}
\begin{equation}
S^{(4D)} =  \frac18\int_{\mathcal{M}_{4}}\!\ve_{ABCDE}\Phi^{A}\mathbb{F}^{BC}\mathbb{F}^{DE},\label{eq:phi-FF-action}
\end{equation}
with the \ADS$_5$ curvature
\begin{equation}
\mathbb{F}^{AB}= d\mathbb{A}^{AB}+\mathbb{A}^{A}\!_{C}\mathbb{A}^{CB},
\end{equation}
which in terms of the SO(1,3) components reads  
\begin{align*}
\mathbb{F}^{IJ} & =R^{IJ}-\la^2e^{I}e^{J},\quad(R^{IJ}=d\omega^{IJ}+\omega^{I}\!_{K}\omega^{KJ})\\
\mathbb{\mathbb{F}}^{I4} & = \la D e^{I}.\quad\quad\quad\quad\quad(D e^{I}=d e^{I}+\omega^{I}\!_{J}e^{J})
\end{align*}
The infinitesimal \ADS$_5$ gauge transformations which leave 
invariant the Chamseddine action may be written as
\eq \ba{l}
\d  \mathbb{A}^{AB} =  d \e^{AB} 
+ \mathbb{A}^A{}_C \,\e^{CB}-\mathbb{A}^B{}_C\, \e^{CA}\,,\esp
\d\F^A = \F_B\,\ve^{BA}\,.
\ea\eqn{ADS5-transf}  
The equations of motion from \equ{eq:phi-FF-action} are
\begin{align}
\dfud{S^{(4D)} }{\Phi^{A}} = \frac18\ve_{ABCDE}\mathbb{F}^{BC}\mathbb{F}^{DE} & =0,\label{eq:so5-eom1}\\
\dfud{S^{(4D)} }{A^{AB}} = \half\ve_{ABCDE}\mathbb{D}\Phi^{C}\mathbb{F}^{DE} & =0\,.\label{eq:so5-eom2}
\end{align} 
Or, in terms of the SO(1,3) components
\eq\ba{l}
\dfud{S}{e^I} = -\dfrac{\la}{2}\ve_{IJKL}(D\Phi^{J}+s\la e^{J}\Phi^{4})
(R^{KL}-s\la^2e^{K}e^{L})  = 0\,,\esp
\dfud{S}{\om^{IJ}}  = \half\ve_{IJKL}\lp(d\Phi^{4}-\la e_{I'}\Phi^{I'})(R^{KL}-s\la^2e^{K}e^{L})
+\la(D\Phi^{K}+s\la e^{K}\Phi^{4})D e^{L}\rp =0\,,\esp
\dfud{S}{\F^4} = \frac18\ve_{IJKL}(R^{IJ}-s\la^2e^{I}e^{J})(R^{KL}
-s\la^2e^{K}e^{L})  =0\,,\esp
\dfud{S}{\F^I} = \dfrac{\la}{2}\ve_{IJKL}D e^{J}(R^{KL}-s\la^2e^{K}e^{L})  =0\,.
\ea\eqn{eq:auxiliar}
\subsubsection{Introducing matter -- Continuity equation}\label{introducing matter}

We may introduce matter adding to the action \equ{eq:phi-FF-action}
a matter term $S_{\rm m}$ which we will suppose \ADS$_5$ gauge invariant
and independent of the scalar field $\F^A$. 
\ADS$_5$ gauge invariance of the total action
\eq
S=S^{(4D)}[e,\om,\F] +  S_{\rm m}[e,\om]\,,
\eqn{total-action}
can be expressed through the 
local ``Ward identity'' 
\[
W_{AB} S := -\mathbb{D}\dfud{S}{\mathbb{A}} + \F_A\dfud{S}{\F^B} -
\F_B\dfud{S}{\F^A} = 0\,.
\]
We shall be interested in particular to the Ward identity linked to invariance along 
the generators $M_{I4}$:
\eq
W_I S := - \la e^J\dfud{S}{\om^{IJ}} 
- \dfrac{1}{\la} D\dfud{S}{e^{I}} +\F_{I}\dfud{S}{\F^{4}} - s\F^{4}\dfud{S}{\F^{I}} = 0\,.
\eqn{WI}
Note that these identities hold separately for both actions $S^{(4D)}$ and $S_{\rm m}$.
Defining 
\[
\TT_I: = \dfud{S_{\rm m}}{e^I}\,,\quad \TT_{IJ} := \dfud{S_{\rm m}}{\om^{IJ}}\,,
\]
we may rewrite \equ{WI} as
\eq\ba{l}
\la e^J\TT_{IJ} +\dfrac{1}{\la} D\TT_I = \esp
\quad -\la e^J\dfud{S^{(4D)}}{\om^{IJ}} 
-\dfrac{1}{\la}D \dfud{S^{(4D)}}{e^{I}} +\F_{I}\dfud{S^{(4D)}}{\F^{4}} 
- s\F^{4}\dfud{S^{(4D)}}{\F^{I}} = W_I S^{(4D)} = 0\,,
\ea\eqn{inv-Cham}
the last equality expressing the invariance of $S^{(4D)}$. This leads to the general continuity equation
\eq
\laš2 e^J\TT_{IJ} + D\TT_I  = 0\,.
\eqn{general-cont}
The 3-form $\TT_I$ is related to the energy-momentum tensor components
$\TT^N{}_I$  in the tetrad frame by
\eq 
\TT_I = \frac16 \ve_{NJKL}\, \TT^N{}_I e^Je^K e^L\, 
\eqn{energy-mom-tensor}
If $\TT_{IJ}=0$, \ie if the matter action $S_{\rm m}$
does not depend on the spin connection $\om$, \equ{general-cont} is interpreted as the continuity equation for energy and momentum.

\subsubsection{Partial gauge fixing of the Chamseddine theory}

From  the  gauge transformations \equ{ADS5-transf}
leaving the action 
\equ{eq:phi-FF-action} invariant,
one sees that a possible partial gauge fixing is given by the four 
conditions

\eq
\Phi^{I}=0\,,\quad I=0,\cdots,3\,.
\eqn{Phi-g-fix}
The total action, including matter, then reduces to 
\begin{align}
\bar{S} & =\frac18 \int_{\mathcal{M}_{4}}\!
\ve_{IJKL}\Phi^{4}\mathbb{F}^{IJ}\mathbb{F}^{KL} + S_{\rm m} \nonumber\\
 & =\frac18 \int_{\mathcal{M}_{4}}\!\ve_{IJKL}\Phi^{4}(R^{IJ}-s\la^2e^{I}e^{J})
 (R^{KL}-s\la^2e^{K}e^{L}) + S_{\rm m}\,,\label{total-g-fixed-action}
\end{align}
where the matter action $S_{\rm m}$ is supposed to be independent of 
$\F^A$, as above, but will also be assumed not to depend on the spin connection $\om$
from now on.
The field equations derived from the latter action -- to which we may add a matter action
$S_{\rm m}$, supposed to obey the same \ADS$_5$ gauge invariance as the 
pure Chamseddine part -- are
\eq\ba{ll}
\dfud{\bar S}{e^I} = & -\half s\la^2\Phi^{4}\epsilon_{IJKL}(e^{J}R^{KL}
- s\la^2 e^{J} e^{K} e^{L})  +  \TT_I =0\,,\esp
\dfud{\bar S}{\om^{IJ}} = &\half\epsilon_{IJKL}\lp d\Phi^{4}(R^{KL}-s\la^2e^{K}e^{L}) 
+ 2s\la^2 \Phi^{4}e^{K}D e^{L} \rp  =0\,,\esp
\dfud{\bar S}{\F^4} = &\frac18\epsilon_{IJKL}(R^{IJ}-s\la^2e^{I}e^{J})(R^{KL}-s\la^2e^{K}e^{L}) 
 =0\,,
\ea\eqn{PhiFF-field-eq}
where   $T^I:=De^I$ is the torsion and $\TT_I$ the energy-momentum 3-form 
\equ{energy-mom-tensor}.

Comparison of the first of eqs. \equ{PhiFF-field-eq} with the standard Einstein 
equation in the first order formalism,
\eq\ba{l}
\ve_{IJKL}(e^{J}R^{KL}
- \dfrac{\LA}{3}e^{J} e^{K} e^{L})
= -8\pi G \TT_I  \,, \esp
T^I = De^I = 0\,,
\ea\eqn{Einstein-eq}
suggests to identify $3s\la^2$ with the cosmological constant:
\eq
\LA:=3 s \la^2 \,,
\eqn{lambda}
and to define the function 
\eq
G(x) := -\frac{3}{8\pi\LA\Phi^4(x)}\,,
\eqn{Newton constant}
as a variable ``Newton parameter'', proportional to the inverse of the dilation field $\Phi^4$. With this, the field equations 
\equ{PhiFF-field-eq} take the form
\begin{align}
\epsilon_{IJKL}\lp e^{J}R^{KL}
- \dfrac{\LA}{3}e^{J} e^{K} e^{L}\rp & = -8\pi G(x) \TT_I\,,\nonumber\\
\epsilon_{IJKL}\lp dG(x)\lp R^{KL}-\dfrac{\LA}{3}e^{K}e^{L}\rp 
- 2\dfrac{\LA}{3}G(x)e^{K}D e^{L}\rp& =0\,,
\label{definitive-PhiFF-field-eq}\\
\epsilon_{IJKL}\lp R^{IJ}-\dfrac{\LA}{3}e^{I}e^{J}\rp
\lp R^{KL}-\dfrac{\LA}{3}e^{K}e^{L}\rp 
& =0\nonumber\,,
\end{align}
where we have emphasized the $x$-dependence of the Newton parameter $G$.

Let us finish this subsection with some comments:
\begin{enumerate}
\item The theory is clearly singular in $\LA=0$. 
This value would correspond to a vanishing $s$ in the 
\ADS$_5$ metric \equ{metricADS5}, which thus would become singular.
\item
The first field equation in \equ{PhiFF-field-eq} has the form of the usual Einstein equation 
in first order formalism,  but with with varying Newton coupling parameter $G(x)$. 
The second one determines the torsion $T^I$ in terms 
of the basic fields  $\om^{IJ}$,  $e^I$ and $G$. In particular, 
the torsion is zero if $G$ is constant.
The third equation is a new constraint. One must emphasize that the torsion here is not an independent field.
\item
A canonical analysis~\cite{futurework} shows that the 
number of physical degrees of freedom of the theory is 3: this 
corresponds to the two degrees of freedom of the gravitational field plus one 
corresponding to the scalar dilaton field $\Phi^4$ -- the Newton coupling parameter $G$.
\item
In the absence of matter, an obvious trivial solution is the constant curvature
and torsion free (anti-)de Sitter space: $R^{IJ}$ = $\dfrac{\LA}{3}e^{I}e^{J}$. 
\item The last equation, which clearly admits the constant curvature
solution, is also compatible with non-trivial solutions, 
as the examples treated below do show.
\item
It is interesting to note that the gauge fixing condition 
\equ{Phi-g-fix} is nothing 
but the first of the gauge fixing conditions \equ{two-gaugefix} 
(see the definition \equ{eq:SO5-Phi}).
\end{enumerate}

\subsubsection{Energy-momentum continuity}

In Einstein theory, the continuity equation for the energy-momentum 
tensor reads, in the first order formalism used here:
\eq
D\TT_I =0\,,
\eqn{continuity-TT}
where $D$ is the exterior derivative with respect to the 
spin connection $\om^{IJ}$ and the 3-form $\TT_I $ is related 
to the energy momentum tensor by \equ{energy-mom-tensor}. 
The continuity equation 
\equ{continuity-TT} follows from the Einstein field equations\equ{Einstein-eq} 
and the Bianchi identity
$DR^{IJ}=0$. 
 As we saw in the subsection \ref{introducing matter}, 
It turns out that it still holds in our case, as a consequence of the \ADS$_5$ 
invariance expressed by the indentity \equ{inv-Cham} and of the hypothesis we have made 
that the matter action is independent not only from the scalar fields $\F$, 
but also from the spin connection $\om$.

It is interesting to look at the identity \equ{inv-Cham} with the gauge fixing condition $\F^I=0$
being applied. Taking into account the hypothesis that $S_{\rm m}$ only depends on the tetrad $e$,  this leads to the identity
\[
D\TT_I = \dfrac{\LA}{3}e^J \dfud{\bar{S}}{\om^{IJ}}
 +\left.  \sqrt{\frac{|\LA|}{3}} \F^{4}\dfud{S}{\F^{I}} \right\vert_{\F^I=0} \,,
\]
where $\bar{S}$ is the total gauge fixed action \equ{total-g-fixed-action}, 
and $S$ the total action \equ{total-action} before gauge fixing. Since $D\TT_I=0$, the latter identity shows that the equation
\eq
\left.  \dfud{S}{\F^{I}} \right\vert_{\F^I=0} =0\,,
\eqn{missing-eq}
is valid ``on shell'', \ie if the field equations \equ{PhiFF-field-eq}  of the gauge fixed theory
are satisfied. This is just the equation  of the non-gauge fixed theory obtained by varying 
$\Phi^I$, taken at $\F^I=0$.
In fact, the on-shell validity of \equ{missing-eq} can be derived directly from the 
Ward identity \equ{WI} taken at $\F^I=0$, as one can easily check.

\subsubsection{The  field equations of the Chamseddine model as particular field 
equations of the dimensionally reduced 5D Chern-Sim\-ons theory}

One may ask if the equations of motion derived from the truncated
theory, namely the Chamseddine model equations 
(\ref{eq:so5-eom1},\ref{eq:so5-eom2}), together with the truncation 
equations \equ{eq:truncation} and the $\chi$-independence 
conditions \eqref{eq:zero-mod-cond},
are also solutions of the
equations of motion  \equ{CS5fieldeq} of the full original 
\ADS$_6$ Chern-Simons theory. In what follows
we show the answer is positive.

The field equations of the full CS theory reduced in 4 dimension are
given by (\ref{eq:auxiliar-4}, \ref{eq:auxiliar-3}) together with 
\equ{field-eq-Lorentz-comp}. 
After imposing the
truncation \eqref{eq:truncation} together with the
restriction \eqref{eq:zero-mod-cond} and the relabelling 
\equ{eq:SO5-connec}, \equ{eq:SO5-Phi},
the curvature components take the form
\eq\ba{l}
F^{IJ}  =R^{IJ}-s\la^2e^{I}e^{J}\,,\quad 
F^{I4}  =0\,,\quad 
F^{I5}  = \la D e^{I}\,,\quad 
F^{45}  =0\,,\esp
F_{\chi}^{IJ}  =0\,,\quad 
F_{\chi}^{I4} = - \la D\Phi^{I}- s\la^2e^{I}\Phi^{4}\,\quad 
F_{\chi}^{I5} =0\,,\quad 
F_{\chi}^{45} = \la d\Phi^{4}- \la^2e_{I}\Phi^{I}\,.
\ea\eqn{eq:F-trunc}
Inserting the expressions \eqref{eq:F-trunc} in the 
eight equations (\ref{eq:auxiliar-4}, \ref{eq:auxiliar-3}), we obtain four trivial equations $0=0$, and four
non-trivial ones which are identical to those obtained
from the action of the \ADS$_5$ Chamseddine model, 
Eqs. \equ{eq:auxiliar}.
We conclude that the set of solutions of the equations of motion
of the \ADS$_5$ is a particular subset of the solutions of the general
\ADS$_6$ Chern--Simons theory.

It is noteworthy that the four trivial equations are those derived from the CS action 
by varying the four fields destined to truncation. Had we performed any other sort of truncation
in the field equations, 
we would have obtained  more independent equations than what one obtains directly  from the truncated action.

It is also enlightening to see that the effect of the truncation 
when applied
directly to the original \ADS$_6$ connection, leads to
\begin{align}
\hat{A} & =\tfrac{1}{2}\mathbb{A}{}_{\mu}^{AB}M_{AB} d x^{\mu}
+ \Phi^{A}P_{A} d\chi,
\label{truncated-connection}\end{align}
where $\mathbb{A}^{AB}$ and $\Phi^{A}$ were introduced in 
\eqref{eq:SO5-connec}, \eqref{eq:SO5-Phi}. 
We see that the effect of the truncation is to confine
the \ADS$_5$ symmetry into a 4-dimensional connection, whereas the
``translational'' sector of the group is restricted to the $\chi$
dimension. Then it is clear why the truncation, which may seem not obvious
at first sight, results after some simplifications in a 4-dimensional
\ADS$_5$ gauge theory with $\Phi$ a 4-dimensional scalar transforming
as a vector under \ADS$_5$ transformations.

The curvature associated with the truncated connection 
\equ{truncated-connection} is
\begin{equation}
\hat{F}=\tfrac{1}{4}\mathbb{F}_{\mu\nu}^{AB}M_{AB} d x^{\mu} d x^{\nu}+\mathbb{D}_{\mu}\Phi^{A}P_{A} d x^{\mu} d\chi.
\end{equation}

A straightforward calculation shows that by replacing the above result
into \eqref{CS5fieldeq} we obtain \eqref{eq:so5-eom1}, \eqref{eq:so5-eom2},
the field equations of the truncated action \eqref{eq:phi-FF-action}.

\section{Linear approximations}\label{Linear approximations}

In order to investigate the Newtonian limit of the Chamseddine theory or to
look for the presence of wave-like solutions of
the theory in the vacuum, we split the field variables between background
ones, marked with an index $^0$ on the top, and perturbations as follows:
\begin{equation}
\omega^{IJ} = \mathring{\omega}^{IJ} + a^{IJ},\quad 
e^{I} = \mathring{e}^{I} + h^{I},\quad G = \mathring{G} + \phi\,. 
\end{equation}
Up to terms of order higher than one in the perturbation, 
the curvature $R=d\om+\om^2$ and the torsion 
$T^I = De^I$ read 
\[
 R^{IJ} =  \mathring{R}^{IJ} + \mathring{D} a^{IJ}   \,,\quad
 T^I = \mathring{T}^I + \mathring{D} h^I + a^I{}_J\mathring{e}^J\,,
\]
where $\mathring{D}$ is the 
covariant derivative corresponding to the background connection 
$\mathring{\omega}$. The background considered here is a  constant 
curvature de Sitter space-time,
solution of
\[
\mathring{R}^{IJ} - \frac{\LA}{3} \mathring{e}^I\mathring{e}^J =0\,,
\] 
hence the expresssion $F^{IJ} =R^{IJ}-\frac{\LA}{3}e^I e^J$ is of first order:
\[
F^{IJ} = f^{IJ} \,\,\,(+\mbox{ orders}>1) \,,\quad
 f^{IJ} = \mathring{D}a^{IJ} 
- \dfrac{\Lambda}{3}(\mathring{e}^{I}h^{J}-\mathring{e}^{J}h^{I}) \,.
\]
We shall also assume that the zero-th order Newton parameter 
$\mathring{G}$ is  
a (non-zero) constant. The second field equation then implies that the zero-th order torsion 
 is vanishing: $\mathring{T}^I=0 $.
The first field equation shows that the energy-momentum 3-form $\TT_I$ 
must be considered as of first order,
and the third equation is identically solved up to and including the 
first order.

The first and second field equations read, at first order:
\eq \ba{l}
\ve_{IJKL}\mathring{e}^{J}
\lp \mathring{D}a^{KL} 
- \dfrac{\Lambda}{3}(\mathring{e}^{K}h^{L}-\mathring{e}^{L}h^{K}) \rp 
 =-8\pi G_0 \TT_I\,,\esp
\ve_{IJKL}G_0\,\mathring{e}^K 
\lp \mathring{D} h^I + a^I{}_J\mathring{e}^J \rp =  0\,,
\ea\eqn{lin-field-eq}
where $G_0=\mathring{G}$  is the 
Newton parameter at zeroth order, interpreted as the actual Newton constant.
The second of these equations implies a vanishing torsion at first order, too:
\[
\mathring{D} h^I + a^I{}_J\mathring{e}^J =0\,.
\]
We are thus left  with the first of equations \equ{lin-field-eq}, where
the first order connection $a^{IJ}$ may be solved in terms of the vierbein 
perturbation components $h^I_\m$ and their derivatives through the null torsion condition.
This is just Einstein General Relativity with cosmological constant at first 
order of perturbation, in a de Sitter background.

A first implication is that the theory admits a Newtonian 
limit like Einstein's does.
A second implication concerns the theory with cosmological 
constant in the vacuum. Since at first order the theory coincides 
with Einstein's, we can rely on the results of an extensive 
study made by the authors of~\cite{bernabeu}, where 
they show that, beyond the 
constant curvature solution, there are propagating wave solutions.
We refer to their paper for more details.

\section{Cosmological solutions}\label{Cosmological solutions}

In order to explore the physical content of the Chamseddine model, we look 
in this Section for solutions of the cosmological type and compare them with the 
known $\LA$CDM results~\cite{Planck-data}.

\subsection{Isotropy and homogeneity}
We examine the solutions of the field equations (\ref{PhiFF-field-eq})
considering a space-time foliated by a family of isotropic and homogeneous 
3-dimensional spatial slices, as described by the standard Big Bang cosmology. 
The metric that describes this is the Friedmann-Lema\^itre-Robertson-Walker 
(FLRW) metric, given by
\begin{eqnarray*}
ds^2=-dt^2+a^2(t)\lc \frac{dr^2}{1-kr^2}
+r^2d\theta^2+r^2\sin^2\theta\, d\vf^2\rc\,,
\end{eqnarray*}
depending on the time dependent scale factor $a(t)$ and the space curvature parameter $k$ = $0,\,\pm1$.
The space-time coordinates are the time coordinate $t$ and the spatial 
spherical coordinates $r,\theta,\vf$. 
The FLRW metric admits six isometries generated by six global Killing vectors 
associated with three spatial translation $\xi_{(a)}$ and three 
rotation $\xi_{[ab]}$ invariances -- \ie such that 
$\LL_{\xi_{[ab]}} g_{\mu\nu}$ = $\LL_{\xi_{(a)}} g_{\mu\nu}$ = 0~-- which read in Cartesian coordinates  $x^a,\,a=1,2,3$:
\begin{eqnarray*}
\xi_{(a)}=\sqrt{1-kr^2}\partial_a, \hspace{20pt} \mbox{and} \hspace{20pt} \xi_{[ab]}=x_a\partial_b-x_b\partial_a.
\end{eqnarray*}
We assume that the torsion and the scalar field (the Newton parameter $G$) have 
the same isometries as the metric, \ie 
$\LL_\xi T^\rho\!_{\mu\nu}=0$ 
and $\LL_\xi G=0$. These conditions imply $G=G(t)$, 
and the nonvanishing components of $T^{\alpha}\!_{\mu\nu}$ 
are\footnote{Torsion $T^I{}_{\m\n}$ is defined by 
$T^I=De^I$, whereas $T^\rho{}_{\m\n}
= e^\rho_I T^I{}_{\m\n}$.}~\cite{Tolosa-Zanelli}
 \begin{eqnarray*}
 &&T^{r}\!_{\theta\varphi}=
 2f(t)a(t)r^2\sqrt{1-kr^2}\sin\theta, \hspace{20pt}  
T^{\varphi}\!_{r\theta}=\frac{2f(t)a(t)}{\sqrt{1-kr^2}\sin\theta},\\
  && T^{\theta}\!_{r\varphi}=
  -\frac{2f(t)a(t)\sin\theta}{\sqrt{1-kr^2}}, \hspace{70pt} 
  T^{r}\!_{rt}=T^{\theta}\!_{\theta t}=T^{\varphi}\!_{\varphi t}=h(t)
 \end{eqnarray*}
where $f(t)$ and $h(t)$ are functions of time to be determined by the field 
equations.

Working in the first order formalism, we have to choose a 
corresponding pa\-ra\-me\-tri\-za\-tion of the vierbein. A convenient 
choice~\cite{Tolosa-Zanelli} is\footnote{This choice amounts to a gauge fixing of the local Lorentz invariance.}:
\begin{eqnarray*}
&&e^0=dt, \hspace{50 pt} e^1=\frac{a(t)}{\sqrt{1-kr^2}}dr,\\
&&e^2=a(t)rd\theta\,, \hspace{23 pt} e^3=a(t)r\sin\theta d\varphi\,.
\end{eqnarray*}
In this basis the torsion 2-form becomes
\[
T^0=0\,,\quad
T^i=h(t)e^ie^0+f(t)\ve^{i}\!_{jk}e^je^k\,.
\]
(The indices $i,j\cdots$ take the values 1,2,3.)
The spin connection $\om$ which gives rise to this torsion reads
\begin{eqnarray*}
&&\omega^{0i}=(H+h)e^i, \hspace{60 pt} \omega^{12}=-\frac{\sqrt{1-kr^2}}{ar}e^2-fe^3,\\
&&\omega^{31}=\frac{\sqrt{1-kr^2}}{ar}e^3-fe^2, \hspace{20 pt} \omega^{23}=-\frac{\mbox{cot}\theta}{ar}e^3-fe^1, 
\end{eqnarray*}
where 
\eq
H:=\dot{a}(t)/a(t)
\eqn{def-Hubble}
is the Hubble parameter.
The Riemann curvature is given by 
\begin{eqnarray*}
&&R^{0i}=\lp (\dot{H}+\dot{h})+H(H+h)\rp e^0 e^i+f(H+h)\ve^{i}\!_{jk}e^j e^k,\\
&&R^{ij}=\lp (H+h)^2+\frac{k}{a^2}-f^2\rp e^i e^j+(\dot{f}+Hf)\ve^{ij}\!_{k}e^k e^0\,.
\end{eqnarray*}
Consequently
\begin{eqnarray*}
&&{F}^{0i}=\lp (\dot{H}+\dot{h})+H(H+h)-\frac{\LA}{3}\rp e^0 e^i+f(H+h)\ve^{i}\!_{jk}e^j e^k,\\
&&{F}^{ij}=\lp (H+h)^2+\frac{k}{a^2}-f^2-\frac{\LA}{3}\rp e^i e^j+(\dot{f}+Hf)\ve^{ij}\!_{k}e^k e^0.
\end{eqnarray*}

\subsection{Field equations}

We assume matter to consist of a perfect fluid of 
density $\rho_{\rm m}$ and 
pressure $p_{\rm m}$, with an energy-momentum tensor 
$
\TT^I{}_J = \mbox{diag } (-\rho_{\rm m},p_{\rm m},p_{\rm m},p_{\rm m})\,.
$
Substituting in the field equations (\ref{PhiFF-field-eq}), 
with $dG=\dot{G} e^0$, we get 
the system of differential equations
\begin{eqnarray}
\label{ce1}
&&U^2+\frac{k}{a^2}-f^2-\frac{\Lambda}{3}
=\frac{8\pi G}{3}\rho_{\rm m} \,,\\
\label{ce2}
&&  U^2+\frac{k}{a^2}-f^2-\Lambda+2\left(\dot{U}+HU\right)\
= -8\pi G \,p_{\rm m},\\
\label{ce3}
&&\dot{G}\left(U^2+\frac{k}{a^2}-f^2-\frac{\Lambda}{3}\right)
-\frac{2\Lambda}{3}Gh=0,\\
\label{ce4}
&&f\left(\dot{G}U - \frac{\Lambda}{3}G\right)=0,\\
\label{ce5}
&&\left(U^2+\frac{k}{a^2}-f^2-\frac{\Lambda}{3}\right)
\left(\dot{U}+HU-\frac{\Lambda}{3}\right)
-2fU\left(\dot{f}+Hf\right)=0,
\end{eqnarray}
where $U:=H+h$ and $G=G(t)$ is the Newton coupling parameter \equ{Newton constant}.

\subsection{Continuity equations}
A first continuity equation for the energy and pressure of matter 
follows directly from the energy-momentum continuity equation
\equ{continuity-TT}. Calculating the components of the energy-momentum
3-form $\TT_I$, from (\ref{energy-mom-tensor}) one finds
 \begin{eqnarray*}
 \TT_0&=&-\frac{\rho_{m}(t)}{6}\epsilon_{ijk}e^{i}e^{j}e^{k},\\
 \TT_i&=&-\frac{p_{m}(t)}{6}\epsilon_{ijk}e^{0}e^{j}e^{k},\\
 \end{eqnarray*}
consequently
\begin{eqnarray*}
 \TT_0&=&-\rho_{m}(t)e^{1}e^{2}e^{3}=-\frac{\rho_{m}(t)a(t)^{3}r^{2}\sin\theta}{\sqrt{1-kr^{2}}}dr\wedge d\theta\wedge d\vf,\\
 \TT_1&=&-p_{m}(t)e^{0}e^{2}e^{3}=-p_{m}(t)a(t)^{2}r^{2}\sin\theta dt\wedge d\theta\wedge d\vf,\\
 \TT_2&=&-p_{m}(t)e^{0}e^{3}e^{1}=-p_{m}(t)\frac{a(t)^{2}r\sin\theta}{\sqrt{1-kr^{2}}} dt\wedge d\vf \wedge dr,\\
 \TT_3&=&-p_{m}(t)e^{0}e^{1}e^{2}=-p_{m}(t)\frac{a(t)^{2}r}{\sqrt{1-kr^{2}}} dt\wedge dr \wedge d\theta,\\
 \end{eqnarray*}
The equation $D\TT_0=0$ yields the density-pressure-torsion
continuity equation
\eq
\dot\rho_{\rm m} +3H(p_{\rm m}+\rho_{\rm m}) +3h\,p_{\rm m} = 0\,.
\eqn{cont-rho-p-h}
The equations $D\TT_i=0$ for $i=1,2,3$ are trivially satisfied, 
being of the form $0=0$.

Note the torsion dependence in the last term of  \equ{cont-rho-p-h}. However, for
a matter with zero pressure (cold matter, dust), this continuity equation 
takes the usual form~\cite{Padmanabhan}:
\eq 
\dfrac{d}{dt}\lp \rho_{\rm m} a^3\rp = 0\,,\quad\mbox{if }p_{\rm m}=0\,.
\eqn{cont-rho}

A second continuity equation can be found in the following way:
One notes that, substituting $U=H+h$ in the equations (\ref{ce1},\ref{ce2})
leads to analogues of the standard Friedmann equations:
 \eq
H^2 = \frac{8\pi G_0}{3}\rho_{\rm tot}\,,\quad
2\dot{H}+3H^2=-8\pi G_0 \,p_{\rm tot}\,,
\eqn{ce11-12}
where $G_0$ is the Newton constant, taken as the present value of $G(t)$, and
$\rho_{\rm tot}$, $p_{\rm tot}$ are the ``total density and pressure''  
\[
\rho_{\rm tot} = 
\dfrac{G}{G_0}\lp \rho_{\rm m} + \rho_k + \rho_T + \rho_\LA\rp \,,
\quad  p_{\rm tot} = 
\dfrac{G}{G_0}\lp p_{\rm m} +p_k + p_T + p_\LA \rp\,,
\]
with
\[\ba{l}
\rho_k = -\dfrac{3}{8\pi G} \dfrac{k}{a^2}\,\quad
\rho_T = \dfrac{3}{8\pi G}(f^2 - 2Hh - h^2)\,,\quad
\rho_\LA = \dfrac{\LA}{8\pi G}\,\esp
p_k=- \rho_k/3\,,\quad
p_T=\dfrac{1}{8\pi G}(2\dot{h}+4Hh+h^2-f^2)\,,\quad
p_\LA = -\rho_\LA\,.
\ea\]
$\rho_T$ and $p_T$ may be interpreted as the contributions of the torsion
to the total density and pressure $\rho_{\rm tot}$ and $p_{\rm tot}$.
As a consequence of the Friedmann-like equations \equ{ce11-12}, 
the total density and pressure satisfy the continuity equation
\begin{eqnarray*}
\dot{\rho}_{\rm tot}+3H(\rho_{\rm tot}+p_{\rm tot})=0\,.
\end{eqnarray*}

\subsection{Pressure-less matter with $\LA>0$ and $k=0$}
In this subsection we present the general solution of the equations
\equ{ce1}-\equ{ce5} in the case of pressure-less 
matter (cold matter or dust), with $p_{\rm m}=0$,  with a positive 
cosmological constant $\LA$ and a null curvature parameter $k$,
as favoured by the observational results~\cite{Planck-data}. 
From the equation \equ{ce4} follows 
\eq
\mbox{either}\quad f(t)=0\,,\quad \mbox{or} \quad
\dot{G} U-\frac{\Lambda}{3}G=0\,.
\eqn{ce4'}
We have first checked that the former case leads to the "trivial" solution
of a null torsion de Sitter space with cosmological constant $\LA$,
the vierbein or the metric being defined by the scale parameter
$a(t) = \exp(\sqrt{\LA/3}t)$.

We hence assume the function $f(t)$ to be non-vanishing. The equations
to be solved are the equations  (\ref{ce1}-\ref{ce3}), \equ{ce5} and
the second of 
\equ{ce4'}, together with the Hubble parameter definition
 \equ{def-Hubble} in terms of the scale $a(t)$. The general solution 
 is given by the following 
 expressions\footnote{\label{mathematica}Solution 
 obtained using the program 
Mathematica~\cite{Mathematica}.}, where the time coordinate 
has been redefined by
\[
\tau(t):=\sqrt{\frac{\LA}{3}}\,t\,.
\]
{\it Scale parameter:}
\eq\ba{l}
a(t)=C_4\lp 3e^{\tau(t)}+ C_3 e^{-\tau(t)}\rp^{1/3}
\lp \cosh\left(\tau(t)-C_1\right)\rp^{2/3}
\ea \eqn{scale-par}
{\it Torsion parameter $f(t)$:}
\eq\ba{l}
f(t) =\dfrac{\sqrt{\LA}}{3}
\lc \lp 
-9 e^{2\tau(t)}-3 C_3 + 
\lp 6e^{2\tau(t)}-2 C_3\rp\tanh(\tau(t)-C_1) \right.\right.\\
\phantom{f(t) =} \left.\left.+\lp 3e^{2\tau(t)}+ C_3\rp
\tanh^2(\tau(t)-C_1)
\rp 
\big{/}\lp 3e^{2\tau(t)}+ C_3\rp
\rc^{1/2}\,.
\ea\eqn{Torsion-f}
{\it Torsion parameter $h(t)$:}
\eq\ba{l}
h(t) =   \sqrt{\dfrac{\LA}{3}}
\dfrac{\lp 
-3 e^{2\tau(t)}+ C_3 + 
\lp 3e^{2\tau(t)}+ C_3\rp\tanh(\tau(t)-C_1)
\rp} 
{ 9e^{2\tau(t)}+3 C_3}\,.
\ea\eqn{Torsion-h}
{\it Hubble parameter $H=\dot a/a$:}
\eq
H(t) = \sqrt{\dfrac{\LA}{3}} \tanh(\tau(t)-C_1) - h[t]\,.
\eqn{Hubble-par}
{\it Newton parameter $G(t)=-3/(8\pi\LA\F(t))$  (c.f. \equ{Newton constant}:}
\eq
G(t) = C2 \sinh(\tau(t)-C_1)  \,.
\eqn{Newton-par}
{\it Cold matter density:}
\eq\ba{l}
\rho_{\rm m}(t) = \dfrac{3}{8\pi G(t)}\lp (H(t)+h(t))^2
-f^2(t)-\dfrac{\LA}{3} \rp\,.
\ea\eqn{CM-density}

The four integration constants $C_1,\,C_2,\,C_3,\,C_4$ and the 
cosmological constant $\LA$ have to be determined by five physical conditions, 
which we choose to be:
\eq\ba{ll}
a(0) = 0\,:\quad& \mbox{hypothesis of a Big Bang}\,,\\
a(t_0) = 1\,:\quad& t_0 = \mbox{present age of the Universe}\,,\\
H(t_0) = H_0\,:\quad& \mbox{present value of the Hubble parameter}\,,\\
G(t_0 = G_0\,:\,:\quad& \mbox{present value of the Newton parameter}\,,\\
\rho_{\rm m}(t_0)=\rho_0\,:\quad& \mbox{present value of the cold matter mass density}\,,
\ea\eqn{phys-conditions}
with the present observational~\cite{Planck-data} and experimental data
 given by
\[\ba{l}
t_0 = 13.8\times 10^9\,\mbox{Gy}\quad \mbox{(1Gy}=10^9 \,\mbox{years}
= 3.1558\time 10^{16}\,\mbox{s)}\,,\\
H_0 =  0.0693 \,\mbox{Gy}^{-1}\,,\\
\rho_0 = 2.664\times 10^{-27}  \mbox{Kg}\,\mbox{m}^{-3}\,,\\
G_0 = 6.674\times 10^{-11}\,\mbox{m}^3 \,\mbox{s}^{-2}\,\mbox{kg}^{-1}\,.
\ea\]

For comparison with the standard $\LA$CDM results, we need the 
$\LA$CDM formula for the scale parameter $a(t)$, for a Universe dominated by cold dark matter of present relative density~\cite{Planck-data}
 $\Omega_{\rm{m}}$ =
$0.309$. With the contribution of radiation neglected, the normalized
$\LA$CDM scale parameter reads~\cite{Padmanabhan}
\eq
a_{\LA{\rm CDM}}(t)=
\lp
\sinh\lp\frac32 H_0 \sqrt{1-\Omega_{\rm{m}0}}\, t \rp
\Big/\sinh\lp\frac32 H_0 \sqrt{1-\Omega_{\rm{m}0}}\, t_0 \rp
\rp^{2/3}
\eqn{LCDM-scale}
\begin{figure}[htb]
\centering
\includegraphics[scale=0.38]{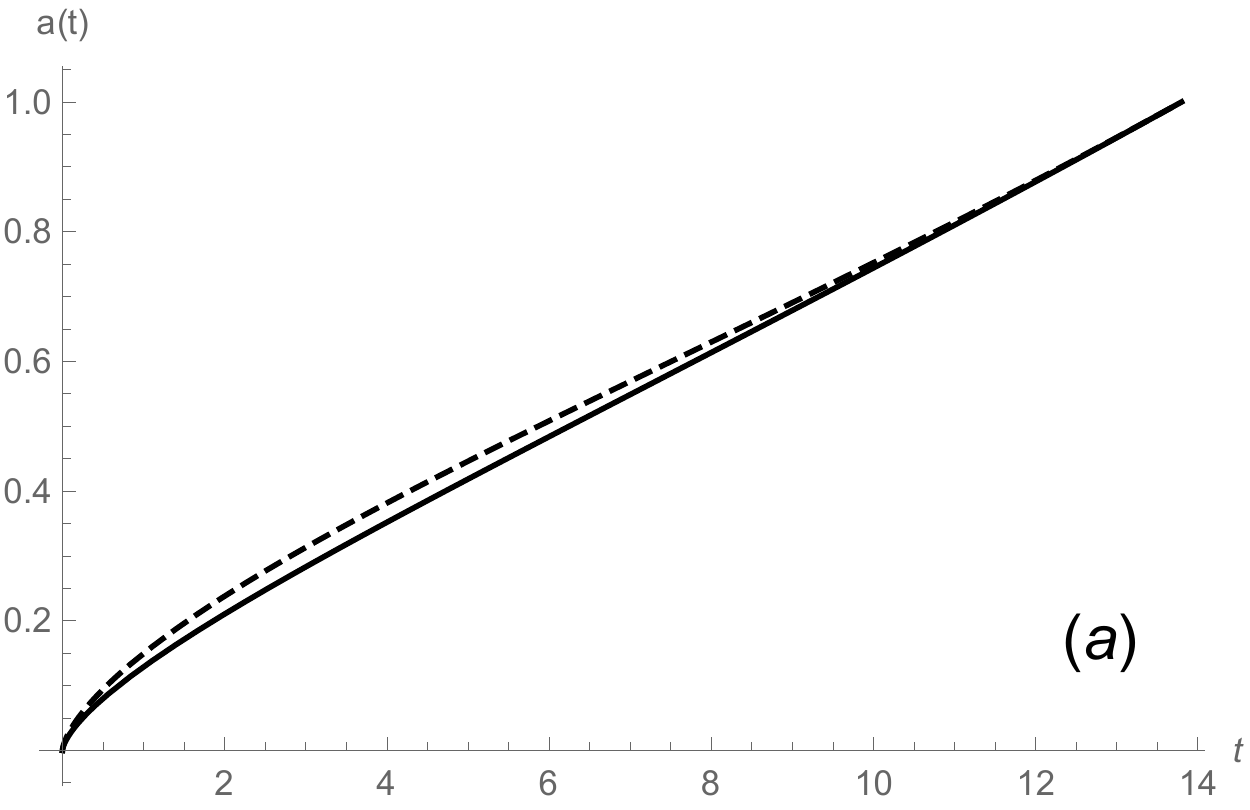}
\includegraphics[scale=0.38]{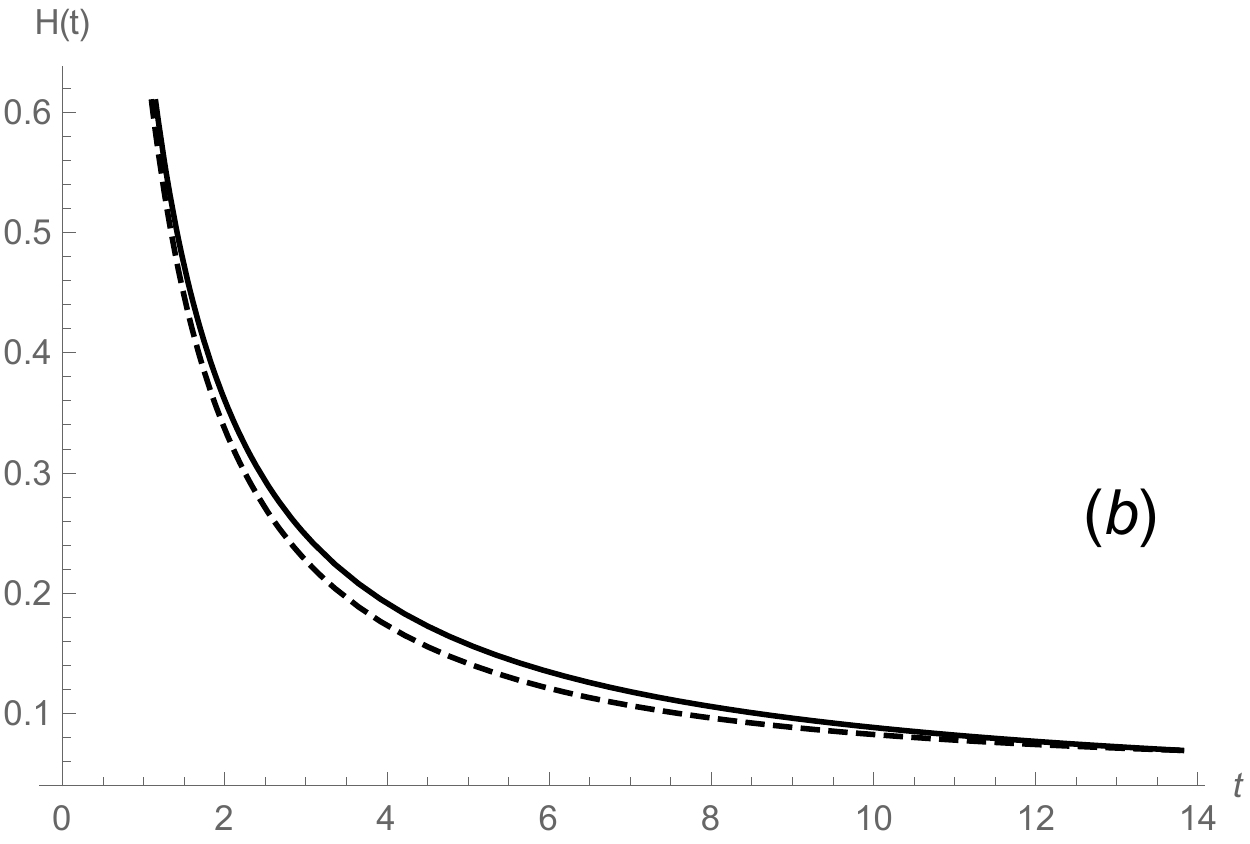}
\includegraphics[scale=0.38]{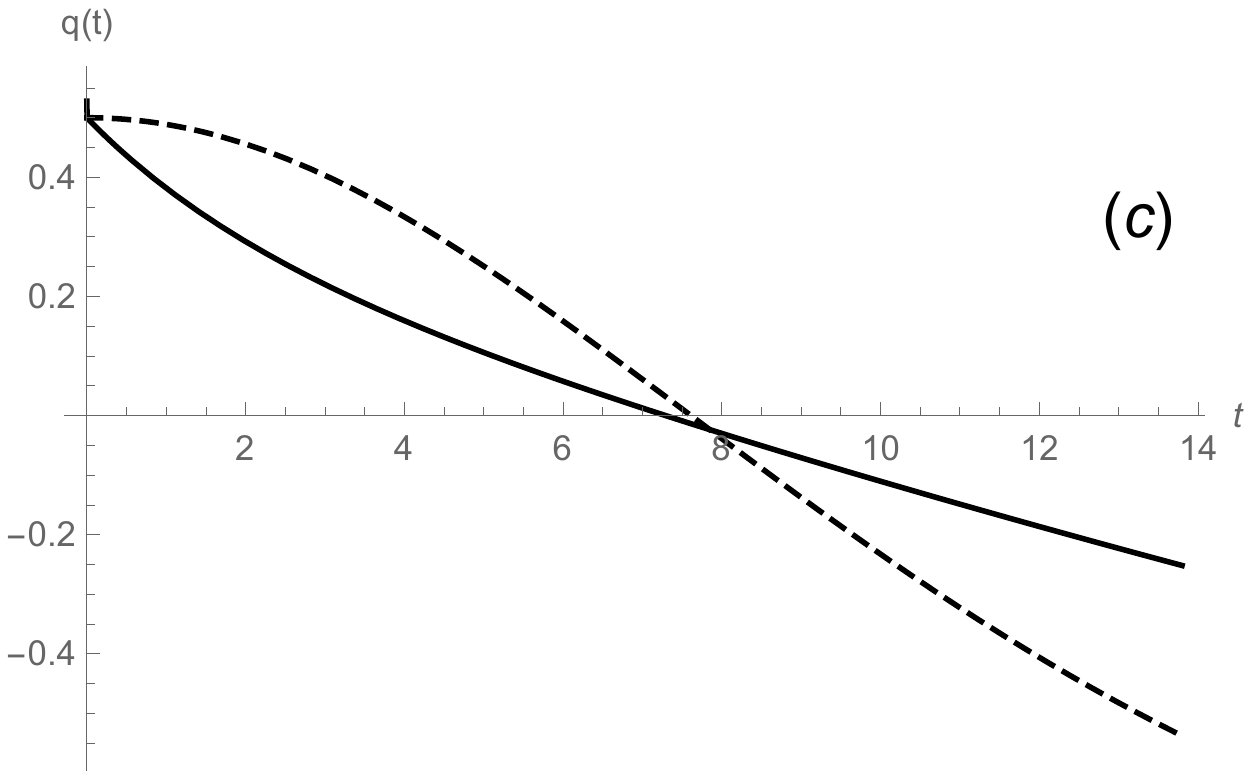}\\[10mm]

\includegraphics[scale=0.38]{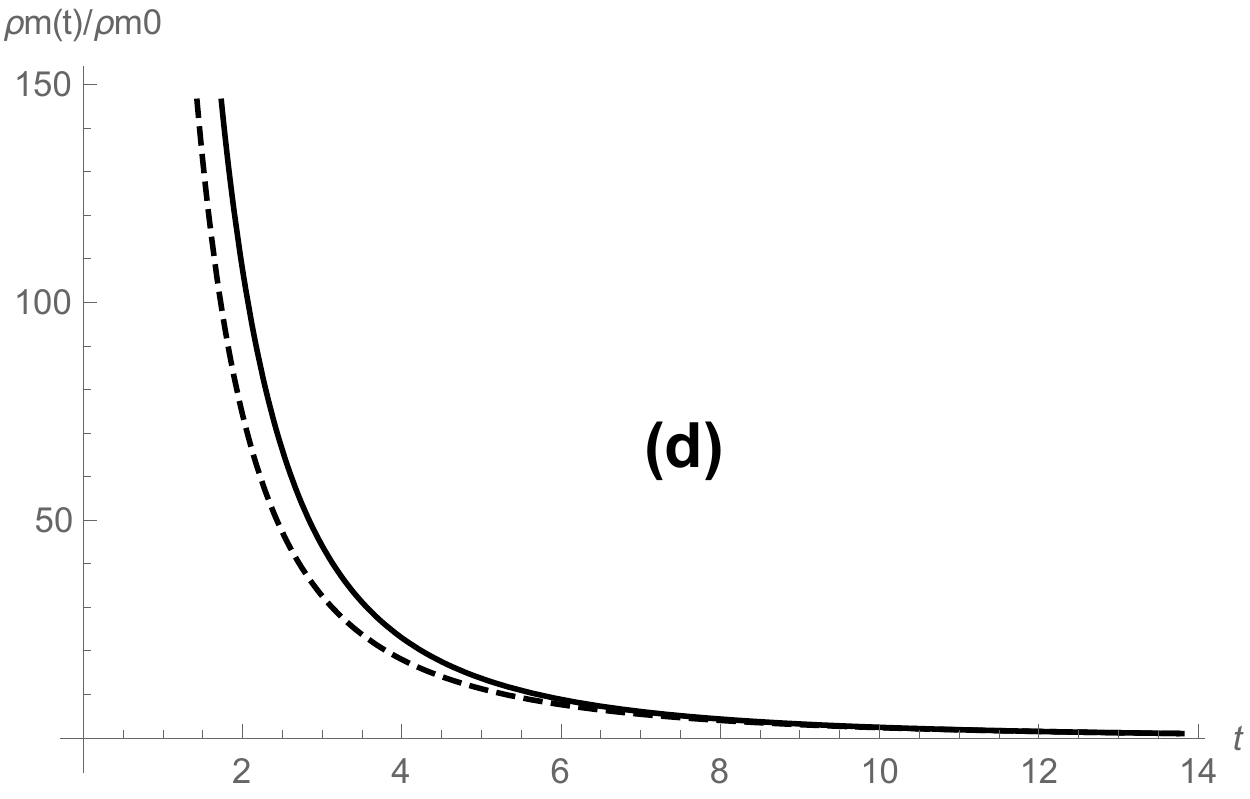}
\includegraphics[scale=0.38]{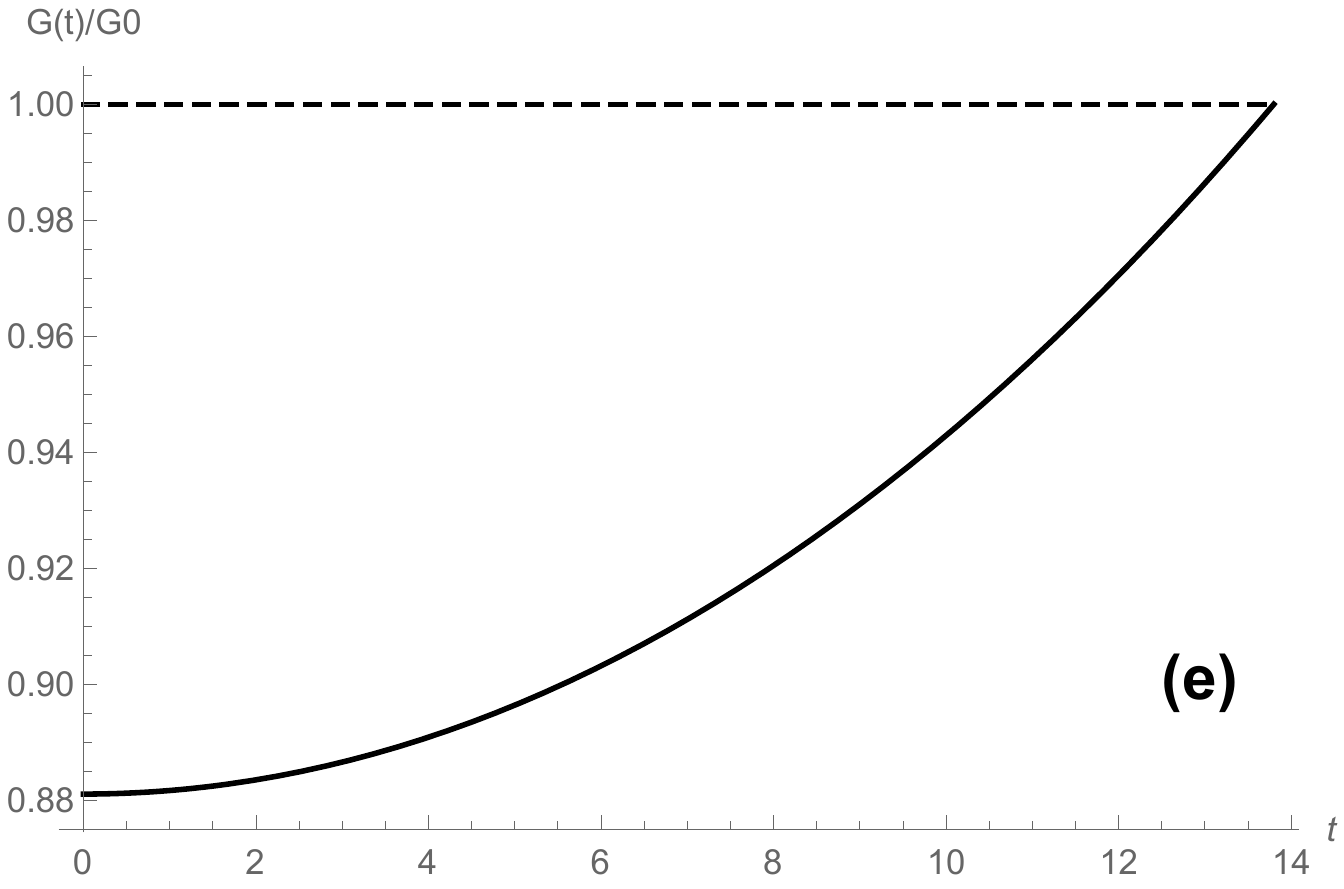}
\caption{\it\small (a) Normalized scale parameter $a(t)$; 
(b) Hubble parameter $H(t)$;
(c) deceleration parameter $q(t)$;
(d) cold matter density $\rho_{\rm m}(t)$;
(e) time-dependent gravitation coupling parameter $G(t)$;
Solid lines: model predictions; dashed lines: 
standard $\Lambda$CDM results. }\label{fig1}
\end{figure}
Fig. \ref{fig1} shows the time evolution of the scale parameter $a$, 
of the Hubble constant $H$, of the deceleration parameter 
$q$ = $-\ddot{a}a/(\dot{a}a^2)$,
of the mass density $\rho_{\rm m}$ and of the normalized Newton parameter 
$G/G_0$, each one being compared with the corresponding $\LA$CDM quantity. 
Excepted for the deceleration $q$, the deviations are rather small. the Newton parameter, which has  to be equal to the actual Newton constant $G_0$ at the present time, shows a slight decrease towards the past, 
achieving $\sim 85\%$ of its present value near of the Big Bang. 
The deceleration $q$ differs notably from the $\LA$CDM one, 
but the time of
the transition between the deceleration and the acceleration era 
almost coincide. The present value $q(t_0)$ = $-0.25$ is 
however only half of the $\LA$CDM value.
\begin{figure}[htb]
\centering
\includegraphics[scale=0.38]{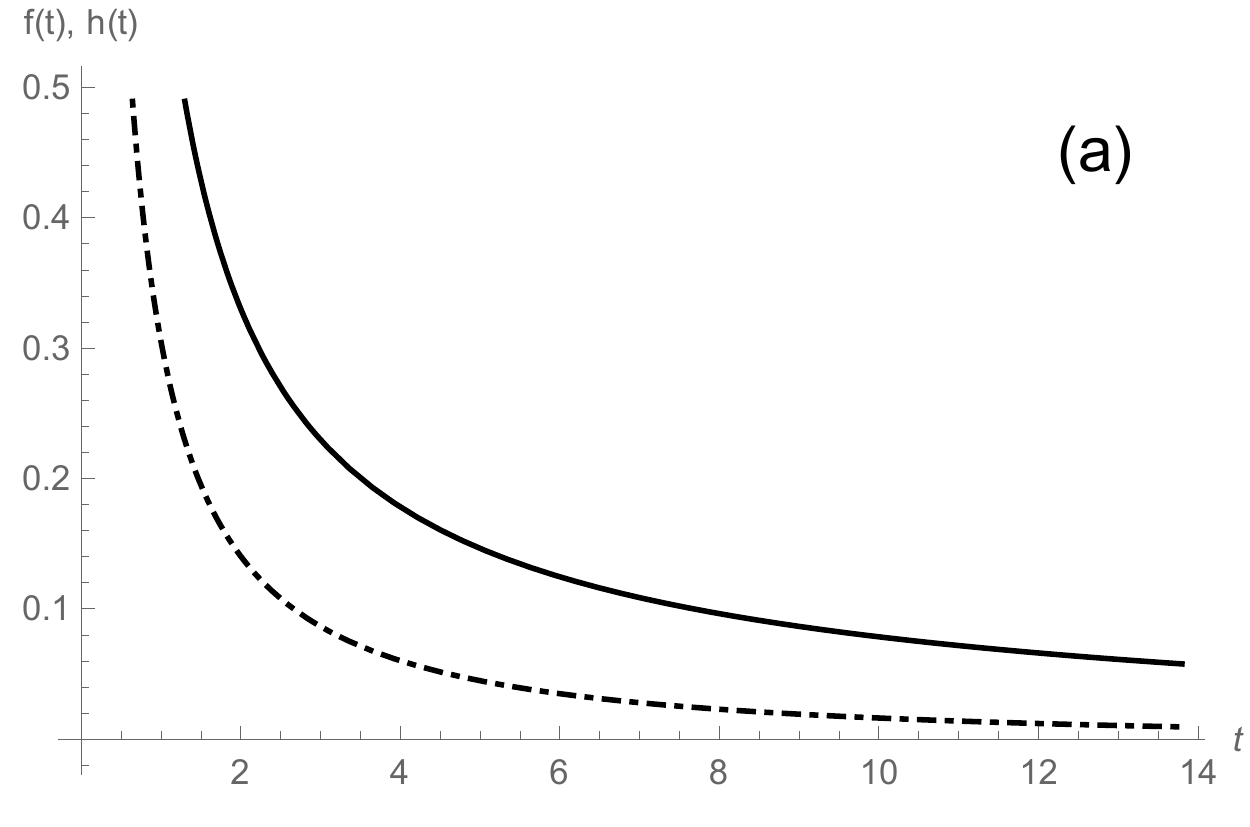}
\includegraphics[scale=0.38]{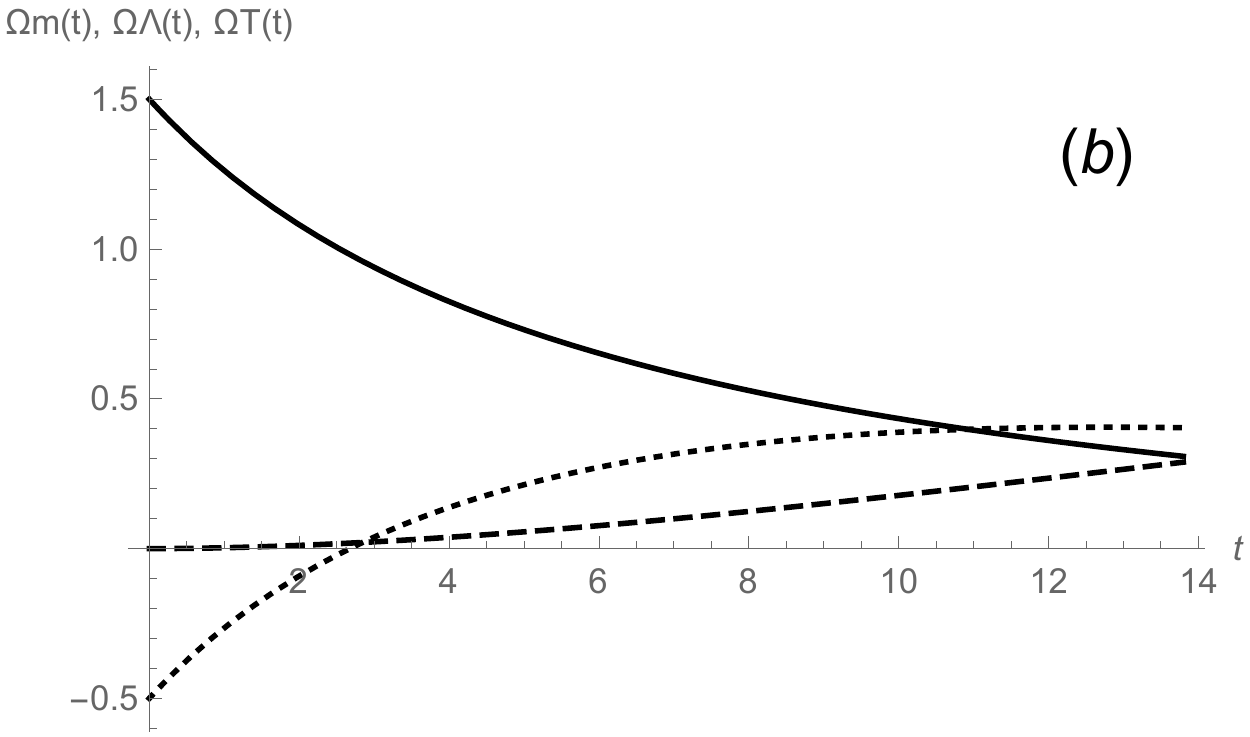}
\includegraphics[scale=0.40]{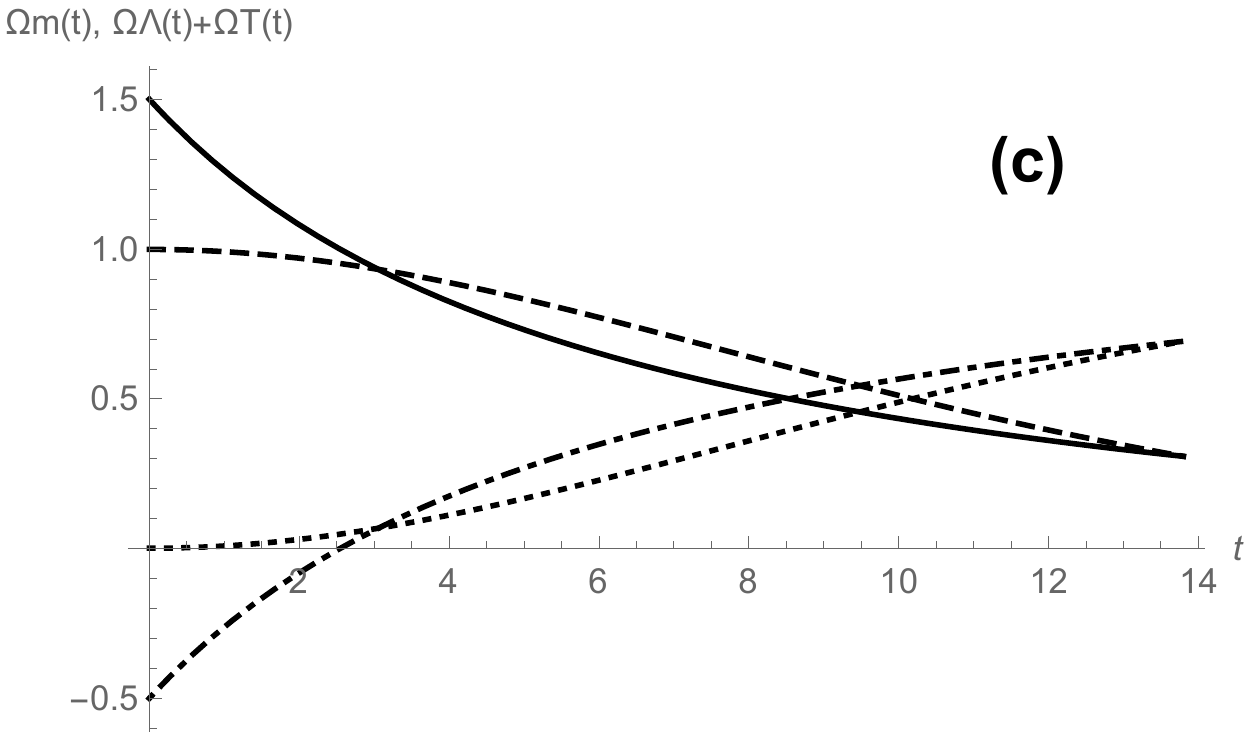}
\caption{\it\small(a) Torsion parameters $f(t)$ (solid line) and $h(t)$ (dashed line); 
(b) Relative densities $\Omega_{\rm m}(t)$ (solid line), $\Omega_\LA(t)$ (dashed line) and  $\Omega_T(t)$ (dotted line);
(c)  $\Omega_{\rm m}(t)$ (solid line)and $\Omega_\LA(t)$ + $\Omega_T(t)$ 
(dot-dashed line); $\LA$CDM results are shown for $\Omega_{\rm m}$ (dashed line) and $\Omega_\LA$ (dotted line).}
\label{fig2}
\end{figure}
Time evolutions of 
the torsion parameters $h$ and $f$, as well as the relative densities
 $\Omega_{\rm m}(t)$, $\Omega_\LA(t)$ and $\Omega_T(t)$ 
for  matter, cosmological constant and torsion, respectively,
are shown in Fig. \ref{fig2}(a-b). 

We observe from
Fig. \ref{fig2}(c) 
that the end of  the cold matter dominance area,  
at $t\sim 10.2$ Gy for $\LA$CDM, occurs
at $t\sim 8.5$ Gy  for our model, matter dominance being 
defined, in the latter case, as the dominance of $\OM_{\rm m}$ over the 
sum $\Omega_\LA$ + $\Omega_T$.

Finally, the present values of the concentrations are:

$\Omega_{\rm m}(t_0)= 0.308  $  (which belongs to the input data),\\
\indent$\Omega_\LA(t_0)= 0.289 $,\\
\indent$\Omega_T(t_0)= 0.403 $.

This has to be contrasted with the $\LA$CDM values 
$\Omega_{\rm m}(t_0)= 0.308$ and $\Omega_\LA(t_0)$ = $0.692$: 
in our model the torsion contributes together with the cosmological 
constant to the acceleration.

Finally, as a matter of verification, we have checked that our solution of the field equations 
does satisfy the continuity equation \equ{cont-rho}.
\subsection{Search for other solutions}

Since torsion may contribute to the acceleration, as in the solution
studied above, one could expect solutions 
presenting a positive present acceleration even with a 
negative cosmological constant. This occurs for instance 
for the class of models investigated in~\cite{Tolosa-Zanelli}. 
In our case, we have checked that there is 
no solution with $\LA<0$ and positive acceleration 
fulfilling the physical boundary conditions represented 
by the present values
of the cold matter density and of the Hubble 
and Newton parameters.
  Another class of solutions with a bounce at some time in the past 
  do exist, but none of them is compatible even roughly with physical
   boundary conditions.

\section{Conclusion and outlooks}\label{Conclusion}

We have seen in rather details how the dimensional reduction 
and truncation of 
the \ADS$_6$ Chern-Simons theory in 5D to the 4D Chamseddine
model is working. In particular, we have shown that the field equations of 
the latter form a subset of the field equations of the former,
  which is a non-trivial result. 
Chamseddine's theory involves a scalar dilaton-like field which we 
have interpreted as a varying Newton coupling parameter.
We have explored the solutions of the field 
equations, focusing on two examples. 
In the first one we have shown the existence,
in a linear approximation, of a Newtonian limit and 
of gravitational waves of the same type as 
the ones of standard GR. The Newton parameter is supposed 
to be constant in the zeroth order and turns out to remain
undetermined at first order. The wave solutions confirm 
the canonical result~\cite{futurework} of three degrees of freedom: 
two for the "graviton" 
and one for the Newton parameter field.
The second example is of the FLRW cosmological type. 
We have found a solution
with boundary conditions corresponding to the present 
values of the physical parameters: Newton and Hubble 
parameters, cold matter mass density.
It shows a behaviour fitting rather well that of the standard $\LA$CDM
model,   at least qualitatively.  
The cosmological constant of this solution turns out to be 
positive, however smaller than that of the $\LA$CDM model, 
the torsion contributing substantially to the present 
acceleration of the expansion.
  A similar but different model has been studied by the authors 
of~\cite{Tolosa-Zanelli}. The main difference is that, in their
action, a scalar field appears as a factor only in
the supplementary term, quadratic in the curvature. In our
case, the scalar field appears as a common factor of the whole 
Lagrangian density and, moreover, the term quadratic in the curvature is not independent due to the constraint of the \ADS$_5$ gauge symmetry
SO(1,4) or SO(2,3). 

A study of the full 5D Chern-Simons theory is under way,
with one spatial dimension being compactified~\cite{futurework}. It will allow to explore a larger domain of solutions, this theory 
possessing 13 degrees of freedom as shown 
in~\cite{Banados-etal}.

Concerning the quantization, the  prospect~\cite{futurework} is 
for a Loop Quantization~\cite{Rovelli,Thiemann} of the 5D CS theory.
Indeed, the latter is generic in the sense of the authors 
of~\cite{Banados-etal},
\ie the scalar or ``Hamiltonian'' constraint is a consequence of the 
other constraints, which are easier to solve~\cite{Rovelli,Thiemann}.

\noindent{\bf Acknowledgments:} This work was partially funded by the
Funda\cao\ de Amparo \`a Pesquisa do Estado de Minas Gerais -- 
FAPEMIG, Brazil (O.P.),
the Conselho Nacional de Desenvolvimento Cient\'{\i}fico e
 Tecnol\'{o}gico -- CNPq, Brazil (I.M., Z.O.  and O.P.)
and the Coordena\cao\ de Aperfei\c coamento de Pessoal de N\ii vel Superior --
CAPES, Brazil (I.M. and B.N.).\\  
\indent Zui Oporto wants to specially thank Dr. Juana Centellas Arias and the Instituto Oncol\'ogico Nacional - Caja Petrolera de Salud -- ION-CPS, Bolivia.

\section*{Appendix}
\appendix

\setcounter{section}{1}

\section*{Conventions and notation}
\subsection{Conventions}

4D and 5D space-time indices: $\mu, \cdots=0, \cdots, 3 $ and 
$\a, \cdots=0, \cdots,4 $

\noindent 3D and 4D space indices: $a, \cdots=1, \cdots, 3 $ and 
$m, \cdots=1, \cdots,4 $

The de Sitter or anti-de Sitter groups SO(n,N-n) are collectively denoted by \ADS$_N$.
Their indices and corresponding invariant metrics are denoted by:
\begin{eqnarray} 
\label{metricADS6}&&\mbox{\ADS}_6:\quad  M,N, \cdots=0, \cdots, 5\,,\quad
\eta_{MN}=\mbox{diag}(-1,1,1,1,1,s )\,,\esp
\label{metricADS5}&&\mbox{\ADS}_5:\quad   A,B, \cdots=0, \cdots, 4\,, 
\eta_{AB}=\mbox{diag}(-1,1,1,1,s)\,,
\end{eqnarray} 
where
 $s$ takes  the values $\pm1$ for dS or AdS, respectively.
4D Lorentz SO(1,3) indices are denoted by $I,\cdots$ = $0,\cdots,3)$,
the corresponding  metric being $\eta_{IJ}=\mbox{diag}(-1,1,1,1)$.
   These metrics and their inverses allow to lower and raise the 
   various group indices.

The respective Levi Civita symbols are defined as 
\begin{eqnarray*}
\ve_{MNPQRS}&=&\left\{
	\begin{array}{l}
	 \ve_{012345}:=1\hspace{30pt} \\
	 \ve_{ABCDE4}:=\ve_{ABCDE}
	 \end{array}
	      \right.\\
\ve_{ABCDE}&=&\left\{
	\begin{array}{c}
	 \ve_{01234}:=1\hspace{30pt} \\
	 \ve_{IJKL4}:=\ve_{IJKL}
	 \end{array}
	      \right.\\
\ve_{IJKL}&=&\left\{
	\begin{array}{c}
	 \ve_{0123}:=1\hspace{10pt} \\
	 \ve_{0ijk}:=\ve_{ijk}
	 \end{array}
	      \right.
\end{eqnarray*}
for the internal spaces, and 
\begin{eqnarray*}
\ve^{\a\b\g\d\ve}&=&\left\{
	\begin{array}{l}
	 \ve^{01234}:=1\hspace{10pt} \\
	 \ve^{\m\n\rho\s4}:=\ve^{\m\n\rho\s}
	 \end{array}
	      \right.\\
\ve^{\m\n\rho\s}&=&\left\{
	\begin{array}{c}
	 \ve^{0123}:=1\hspace{10pt} \\
	 \ve^{0abc}:=\ve^{abc}
	 \end{array}
	      \right.
\end{eqnarray*}
for the 5D and 4D space-times.

\subsection{Lie algebra basis}\label{A - Lie algebra basis}
A basis of the Lie algebra \ads$_6$ of the group \ADS$_6$ may be
given by the 15  matrices $M_{PQ}=-M_{QP}$:
\begin{eqnarray*}
(M_{PQ})^{M}{}_{N}:=-(\delta_P^A\eta_{NQ}-\eta_{PN}\delta_Q^M)
\end{eqnarray*}
satisfying the \ads$_6$
 commutation relations
\begin{eqnarray}
\lbrack M_{MN}, M_{PQ}\rbrack = -\eta_{MQ}M_{NP}-\eta_{NP}M_{MQ}+\eta_{MP}M_{NQ}+\eta_{NQ}M_{MP}.
\label{alg-so6}\end{eqnarray}
One can decompose this basis according to representations of the 
5D Lorentz group SO(1,4) as 
\begin{eqnarray*}
M_{MN}=\left\{
	\begin{array}{l}
	 M_{AB}\hspace{50pt} \\
	 P_{A} := \lambda M_{A5}
	 \end{array}
	      \right.
\end{eqnarray*}
where a positive dimensionful parameter $\lambda$  
has been
 introduced, related to a cosmological constant $\LA\sim s\lambda^2$ ($s=\eta_{55}$)
 of a 5D gravitation theory.
The commutation relations read now
\begin{eqnarray}
\lbrack M_{AB}, M_{CD}\rbrack &=& 
-\tilde{\eta}_{AD}M_{BC}-\tilde{\eta}_{BC}M_{AD}
+\tilde{\eta}_{AC}M_{BD}+\tilde{\eta}_{BD}M_{AC}\,,\nonumber\\
\lbrack M_{AB}, P_{C}\rbrack &=& \tilde{\eta}_{AC}P_{B}-\tilde{\eta}_{BC}P_{A}\,,\label{alg-(A)dS_6->Lorentz}\\
\lbrack P_{A}, P_{B}\rbrack &=& s\lambda^2 M_{AB}\,,\nonumber
\end{eqnarray}
with $\tilde\eta_{AB} = \mbox{diag}(-1,1,1,1,1)$.
The ten generators  $M_{AB}$ generate the 5D Lorentz group,
and together with the 5 generators $P_A$, generate the \ADS$_6$ group
for 5D space-timne. The $M_{AB}$ may be represented by the $5\times5$ matrices 
\begin{eqnarray*}
(M_{CD})^{A}\!_{B}:=-(\delta_C^A\tilde\eta_{BD}-\tilde\eta_{CB}\delta_D^A)\,.
\end{eqnarray*}

The first line of \equ{alg-(A)dS_6->Lorentz}, namely
\eq
\lbrack M_{AB}, M_{CD}\rbrack = 
-{\eta}_{AD}M_{BC}-{\eta}_{BC}M_{AD}
+{\eta}_{AC}M_{BD}+{\eta}_{BD}M_{AC}\,,
\eqn{alg-ADS_5}
but this time with the metric
$\eta_{AB} = \mbox{diag}(-1,1,1,1,s)$, gives the commutation 
rules of the Lie algebra of \ADS$_5$. Its decomposition 
according to representations of the 4D Lorentz group reads
\begin{eqnarray*}
M_{AB}=\left\{
	\begin{array}{l}
	 M_{IJ}\hspace{50pt} \\
	 P_{I} := \lambda M_{I4}
	 \end{array}\,.
	      \right.
\end{eqnarray*}
In the same way as above we have introduced the dimensionful parameter
$\lambda$ related  now to the  cosmological constant of a 4D gravitation theory.
Thus
\begin{eqnarray*}
\lbrack M_{IJ}, M_{KL}\rbrack &=& -\eta_{IL}M_{JK}-\eta_{JK}M_{IL}+\eta_{IK}M_{JL}+\eta_{JL}M_{IK}\,,\\
\lbrack M_{IJ}, P_{K}\rbrack &=& \eta_{IK}P_{J}-\eta_{JK}P_{I}\,,\\
\lbrack P_{I}, P_{J}\rbrack &=& s\lambda^2 M_{IJ}\,.
\end{eqnarray*}
We are also interested in the full decomposition of the \ADS$_6$ algebra 
according to representations of the Lorentz group SO(1,3):
\eq
	 M_{IJ}\,,\quad 
	 P_{I} := \lambda M_{I5}\,,\quad 
	 Q_{I} := \lambda M_{I4}\,,\quad R:=M_{45}\,,
\eqn{MPQR}

\eq\ba{l}
\lbrack M_{IJ}, M_{KL}\rbrack = -\eta_{IL}M_{JK}-\eta_{JK}M_{IL}+\eta_{IK}M_{JL}+\eta_{JL}M_{IK}\,,\esp
\lbrack M_{IJ}, P_{K}\rbrack = \eta_{IK}P_{J}-\eta_{JK}P_{I}\,,\quad 
\lbrack M_{IJ}, Q_{K}\rbrack = \eta_{IK}Q_{J}-\eta_{JK}Q_{I}\,,\esp
\lc M_{IJ},R \rc = 0\,,\esp
\lbrack P_{I}, P_{J}\rbrack = s\lambda^2 M_{IJ}\,,\quad
\lbrack Q_{I}, Q_{J}\rbrack = \lambda^2 M_{IJ}\,,\quad
\lbrack P_{I}, Q_{J}\rbrack = \lambda^2 \eta_{IJ} R\,,\esp
\lc P_I,R\rc= s Q_I \,,\quad
\lc Q_I,R\rc= - P_I \,.
\ea\eqn{SO(6)->SO(4)}


\subsection{Dimensions}
The dimensions of the fields and the parameters of the theory,
given in mass units, are:
  \[
 \left|\begin{array}{c|c|c|c|c|c|c|c}
 \hline
  & ds & \omega^{IJ} & e^I  &\lambda  & \Lambda & M_{IJ} & P_{I} \\
 \hline 
 \mbox{dim} & -1 & 1 & 0  & 1 & 2 & 0 & 1\\
 \hline
 \end{array}\right|
\]



\end{document}